\documentclass[twocolumn, prl, reprint, amsfonts, amsmath, amssymb, superscriptaddress, longbibliography, floatfix, aps]{revtex4-2}
\usepackage[T1]{fontenc}
\usepackage{lmodern}
\usepackage{graphicx}
\usepackage[dvipsnames]{xcolor}
\usepackage[colorlinks,linkcolor=Blue,citecolor=Blue]{hyperref}
\usepackage{braket}
\usepackage{mathtools}
\usepackage{enumitem}
\usepackage{microtype}
\usepackage{tikz}

\begin{document}

\renewcommand{\appendixname}{SM Sec.}

\title{Emergent Fermi Surfaces in Periodically Driven Bose Gases}

\author{Li-kun Shi}
\email{likun.shi@zju.edu.cn}
\affiliation{Center for Quantum Matter, School of Physics, Zhejiang University, Hangzhou 310058, China}
\affiliation{Institut f{\"u}r Theoretische Physik, Universit{\"a}t Leipzig, Br{\"u}derstra{\ss}e 16, 04103, Leipzig, Germany}

\author{Oles Matsyshyn}
\affiliation{Division of Physics and Applied Physics, School of Physical and Mathematical Sciences, Nanyang Technological University, Singapore 637371}

\author{Inti Sodemann Villadiego}
\email{sodemann@uni-leipzig.de}
\affiliation{Institut f{\"u}r Theoretische Physik, Universit{\"a}t Leipzig, Br{\"u}derstra{\ss}e 16, 04103, Leipzig, Germany}

\date{\today}

\begin{abstract}
We show that the steady state of a periodically driven Bose gas coupled to a thermal bath features emergent Fermi surfaces both in its high temperature and Bose-Einstein-condensate (BEC) states. These emergent Fermi surfaces are non-analyticities in the occupation function as a function of momentum arising from the lack of detailed balance of collisions away from equilibrium. They are stronger in the BEC than in the normal gas, and can be more robust against finite temperature smearing than the usual equilibrium Fermi surfaces of Fermi gases. We also show that they give rise to analogous rich phenomena, such as Friedel oscillations in the density correlations and quantum oscillations in response to applied magnetic fields controlled by their areas, albeit with marked quantitative differences. Our results establish non-equilibrium conditions as a promising alternative pathway to achieve the long-sought-after emergence of Fermi surfaces in bosonic system. We outline strategies to experimentally discover these exotic states in ultra-cold atomic clouds.
\end{abstract}
\maketitle

\noindent \textit{\color{blue}Introduction.} Fermi gases at zero temperature feature a jump in momentum occupation defining their Fermi surface, which gives rise to a variety of rich phenomena such as their long-ranged Friedel oscillations and their quantum oscillations under magnetic fields. Conversely, an equilibrium ideal Bose gas has no analogous finite-momentum non-analyticity in its occupation function even in the Bose-Einstein condensate (BEC) state~\cite{dalfovo1999theory}. While there is a large body of work investigating the emergence of such Fermi-surface-like non-analyticities in strongly correlated equilibrium bosons (such as 1D bosonic Luttinger liquids~\cite{giamarchi2004quantum,cazalilla2011one}, spin liquids~\cite{savary2017quantum,zhou2017quantum,wen2004quantum,fradkin2013field}, and other exotic bosonic metals \cite{motrunich2007d,lake2021bose}), the question of whether non-equilibrium conditions could facilitate the emergence of Fermi surfaces from bosons has largely remained a blind spot of research. 

In this work, we will demonstrate that even for weakly interacting bosons, non-equilibrium conditions generically lead to steady state occupations with non-analyticities at certain surfaces in momentum space. We will show that these non-equilibrium non-analyticities indeed behave like emergent Fermi surfaces by showing that they give rise to power-law decaying Friedel oscillations and quantum oscillations under magnetic fields with periods controlled by their extremal momentum-space cross-sectional areas. The ingredients to realize these non-equilibrium states are simple: a system of weakly interacting bosons driven out-of-equilibrium by a force oscillating periodically in time, and which is weakly coupled to a simple heat bath. 
\begin{figure}[!htbp]
\centering
\includegraphics[width=\columnwidth]{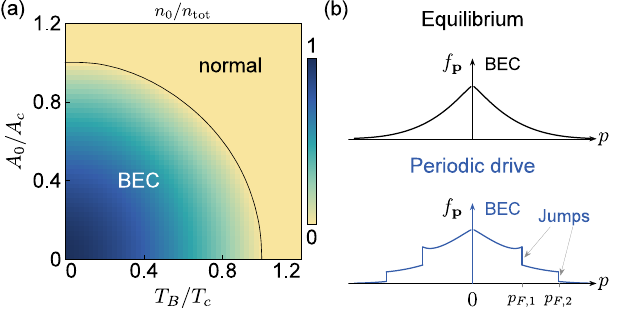}
\caption{(a) Condensate fraction $n_0/n_{\rm tot}$ as a function of bath temperature and drive amplitude. The reference scales are the undriven critical temperature $k_BT_c/\Omega=0.3075$ and the zero-temperature critical drive $A_c/\sqrt{2m\Omega}\simeq0.963$. The smooth curve in (a) guides the boundary of the condensed phase. (b) Schematic occupations in equilibrium (black) and under periodic driving (blue), both with a condensate. The arrows denote the condensates. Periodic driving produces finite-momentum jumps in the excited occupation. Parameters: $n_{\rm tot}/(2m\Omega)^{3/2}=0.01$, $\Delta/\Omega=\sqrt2/12$, and $(c_2\Omega/c_1,\nu_g/(c_1\Omega))=(1,1)$.}
\label{fig1}
\end{figure}
Our work builds on previous studies that showed that periodically driven Fermions display emergent Floquet Fermi surfaces in addition to their equilibrium Fermi surfaces~\cite{matsyshyn2023fermi,shi2024floquet,shi2025ultracritical,matsyshyn2025high}. Remarkably, we will show that such non-equilibrium emergent fermi surfaces can be much stronger for bosons compared to Fermions, specially once they are cooled below their BEC temperature. In fact, we have found that when the bosons are in their BEC state and the bath has a spectral gap, these non-analyticities correspond to true jumps in the boson occupation function that remain sharp at finite temperatures. Therefore, in some cases, these non-analyticities can be more robust than  ordinary equilibrium Fermi surfaces of fermions which inevitably smear out at finite temperature. We have found that these non-analyticities are generically present for a variety of simple particle-conserving  heat baths, but their details depends on their nature. For example, for a BEC in contact with a gapless ohmic bath they become instead  discontinuities of the derivative of the occupation, in contrast to the aforementioned jumps for gapped baths. 

The crucial element behind the emergence of these non-equilibrium Fermi surfaces is the lack of detailed balance in microscopic collisions. The microscopic scattering rates that maintain steady-states are generically highly non-analytic functions, both in equilibrium and non-equilibrium conditions. For example, in the absence of driving, for a bath of harmonic oscillators with a spectral gap $\Delta$, particles in the system with energy $\varepsilon_{\bf p}$ smaller than this gap, are not allowed to decay into lower energy states via emission of quanta into the bath, leading to a non-analytic threshold in the scattering rates at $\varepsilon_{\bf p}=\Delta$. However, in equilibrium, the principle of detailed balance produces a kind of strong ``censorship'' that prevents even such strong non-analyticities of scattering rates from appearing in the occupation functions, and fixes them to have their universal Bose-Einstein or Fermi-Dirac forms. However, as we will show, the periodic driving force that pushes the bosons away from equilibrium lifts this censorship by breaking the detailed balance of scattering rates, and allowing their non-analyticities to be imprinted on the steady-state occupations.

\smallskip

\noindent \textit{\color{blue}Floquet-Boltzmann equation for bosons.} Our system consists of a gas of bosons with a single-particle Hamiltonian $\hat h(t)=[\hat{\bf p}-{\bf A}(t)]^2/(2m)$, driven by a periodic force, $-\partial_t{\bf A}$, with frequency $\Omega$. We use $\hbar=1$ unless explicitly restored. We couple these bosons via their density, $\hat n({\bf r})=\hat\psi^\dagger({\bf r})\hat\psi({\bf r})$, to a large collection of dispersion-less local harmonic oscillators at temperature $T_B$, which act as a bath. The total Hamiltonian is $\hat H(t)=\hat H_S(t)+\hat H_B+\hat H_{SB}$, where $\hat H_S(t)=\int d^3r\,\hat\psi^\dagger\hat h(t)\hat\psi$, $\hat H_B=\int d^3r\sum_\eta\omega_\eta\hat b_\eta^\dagger({\bf r})\hat b_\eta({\bf r})$, and $\hat H_{SB}=\int d^3r\,\hat n({\bf r})\sum_\eta[\chi_0\hat b_\eta({\bf r})+\mathrm{h.c.}]$. Here $\hat b_\eta$ annihilates a bath harmonic mode of frequency $\omega_\eta$ and $\chi_0$ controls the strength of  system-bath coupling. We take these bath modes to form a continuum  with a density of states (DOS), $\nu_B(\omega)$, and will consider different forms for this function to model different kinds of baths. For concreteness we will focus on the case of 3D bosons and circularly polarized driving force, i.e. ${\bf A}(t)=A_0(\cos\Omega t,\sin\Omega t,0)$, but our main qualitatitive finding, namely the existence of non-analytictities in the steady-state occupation, is expected to appear in any dimension and for a broad class of periodic drives~\cite{shi2025ultracritical}.

After a lengthy derivation (detailed in Supplementary Material (SM) Secs.~\ref{FFGR}--\ref{S-condensate-separation}), one can show that in the limit of weak coupling to the bath, $\chi_0\rightarrow0$, the steady-state bosonic single-particle momentum occupation function, $f_{\bf p}$, obeys a Floquet-Boltzmann equation of the form:
\begin{align}
&0=\sum_{\bf q}\big[W_{{\bf q}\to{\bf p}}f_{\bf q}(1+f_{\bf p})-W_{{\bf p}\to{\bf q}}f_{\bf p}(1+f_{\bf q})\big],
\label{eq:boltzmann}
\end{align} 

\noindent with scattering rates given by:
\begin{align}
&W_{{\bf q}\to{\bf p}}=2\pi|\chi_0|^2\sum_{l=-\infty}^{\infty}\Phi^{(l)}_{{\bf q},{\bf p}}S(\varepsilon_{\bf q}-\varepsilon_{\bf p}+l\Omega),
\label{eq:scattering-rates}
\end{align}
where $\varepsilon_{\bf p}={\bf p}^2/(2m)$,  $S(\omega)=[N_b(|\omega|)+\Theta(\omega)]\nu_B(|\omega|)$, $N_b(\omega)=(e^{\beta\omega}-1)^{-1}$, $\beta^{-1}=k_BT_B$, $\Phi^{(l)}_{{\bf q},{\bf p}}=J_l^2[A_0|{\bf q}_\perp-{\bf p}_\perp|/(m\Omega)]$, $J_l$ the Bessel function, and $\perp$ denotes components on the driving $xy$-plane. The function $S(\omega)$ encodes the spectral weight for emission ($\omega>0$) of energy into the bath and absorption ($\omega<0$) from it, assisted by the discrete absorption ($l>0$) and emission ($l<0$) of energy $|l|\Omega$ from the periodic drive, a.k.a. Floquet energy umklapp processes. Similar Floquet kinetic equations have been derived in previous studies \cite{genske2015floquet,seetharam2015controlled,seetharam2019steady,babadi2017theory,dehghani2014dissipative,vorberg2013generalized,vorberg2015nonequilibrium,iadecola2015occupation,esin2018quantized,shi2025ultracritical}; early Floquet master equation treatments appear in Refs.~\cite{blumel1991dynamical,breuer2000quasistationary,hone2009statistical}.

In the limit of vanishing driving amplitude, $A_0\rightarrow 0$, Floquet umklapp processes dissappear, $\Phi^{(l)}_{{\bf q},{\bf p}}\rightarrow\delta_{l,0}$, and scattering rates become $W_{{\bf q}\to{\bf p}}=2\pi|\chi_0|^2 S(\varepsilon_{\bf q}-\varepsilon_{\bf p})$. From this, and using the identity $S(-\omega)=e^{-\beta\omega}S(\omega)$, one can show that in the absence of drive the scattering rates obey the detailed balance condition: $W_{{\bf q}\to{\bf p}}/W_{{\bf p}\to{\bf q}}=e^{\beta(\varepsilon_{\bf q}-\varepsilon_{\bf p})}$. When detailed balance is satisfied the net flow of particle occupation between any pair of states ${\bf q},{\bf p}$ vanishes, namely each pair of terms with fixed ${\bf q},{\bf p}$ cancels inside the sum of Eq.\eqref{eq:boltzmann}, and from this one can show that the steady state is the standard Bose-Einstein function with the temperature of the bath $f_{\bf p}=1/(e^{\beta(\varepsilon_{\bf p}-\mu)}-1)$ \footnote{The chemical potential is determined self-consistently by the total density of bosons, which is conserved by the coupling to the bath.}. However, when the bosons are being periodically driven ($A_0\neq 0$), Floquet umklapp becomes possible and detailed balance is broken. In such case there is no simple universal form describing the steady-state occupation $f_{\bf p}$. 

\smallskip

\noindent \textit{\color{blue}Bosonic Floquet Fermi surfaces.} 
To find the boson non-equilibrium steady-state occupation, $f_{\bf p}$, we have solved numerically Eq.\eqref{eq:boltzmann} as a function of bath temperature $T_B$, and driving amplitude $A_0$, for a variety of bath DOS functions $\nu_B(\omega)=(c_1\omega+c_2\omega^2)\Theta(\omega)+\nu_g\Theta(\omega-\Delta)$, parametrized by constants $c_1,c_2,\nu_g\geq 0$ and $\Delta>0$. We will focus primarily on two types of bath, namely a gapped bath with gap $\Delta$ (realized for $c_1=c_2=0$), and a gapless ohmic bath (realized for $\nu_g=0$), but we will also explore the case of mixed DOS (i.e. when all $c_1,c_2,\nu_g$ are non-zero). As a function of these parameters, we have found that the solutions to Eq.\eqref{eq:boltzmann} always fall into two distinct phases: a non-equilibrium normal state (where all the occupations $f_{\bf p}$ remain finite in the thermodynamic limit), and a non-equilibrium BEC state (where the occupation of the single state ${\bf p}=0$, becomes macroscopic, namely $f_{\bf p=0}/V\rightarrow n_0$ is finite) \footnote{Notice that ${\bf p}$ labels the canonical momentum. The physical mechanical momentum is ${\bf p}-{\bf A}(t)$, and, therefore, the non-equilibrium BEC is in a self-consistent oscillating state in response to the oscillating driving force.}.  A typical phase diagram is shown in Fig.~\ref{fig1}(a) \footnote{Careful manipulations for taking the thermodynamic limit Eq.\eqref{eq:boltzmann} must be performed in order to capture both kinds of solutions, which we detail in SM Secs.~\ref{S-condensate-separation} and \ref{Projected-FBE}.}. 
\begin{figure}[!htbp]
\centering
\includegraphics[width=\columnwidth]{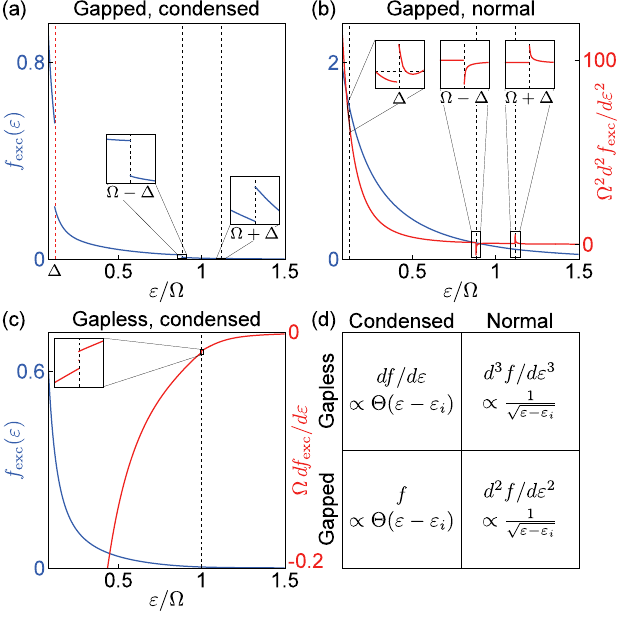}
\caption{Angle-averaged boson occupation $f_{\rm exc}(\varepsilon)$ (blue) of bosons coupled to a gapped bath for: (a) the BEC state at $k_BT_B/\Omega=0.05$, and (b) the normal state at $k_BT_B/\Omega=0.5$, also showing is second derivative (red). (c) Angle-averaged occupation (blue) of a BEC coupled to an ohmic bath ($\nu_g=0$) at $k_BT_B/\Omega=0.05$, and its derivative (red). The driving amplitude in (a--c) is $A_0/\sqrt{2m\Omega}=0.35$, and other parameters are the same as in Fig.~\ref{fig1}(a). (d) Summary of degree of non-analyticities in various states (see SM Secs.~\ref{S-ohmic-nonanalyticities}--\ref{S-surface-hierarchy} for details).}
\label{fig2}
\end{figure}
Remarkably, we have found that both in the normal and BEC driven non-equilibrium steady-states, the occupation $f_{\bf p}$ has sharp non-analyticities at certain spherical surfaces centered around ${\bf p}=0$. This is specially remarkable given that problem is not spherically symmetric: the actual value of the boson occupation, $f_{\bf p}$, changes as a function of the polar angle, $\theta$, relative to the $z$-axis (for details see SM Sec.~\ref{Projected-FBE}). Fig.~\ref{fig2} shows typical solutions displaying these non-analyticities for the spherically averaged occupation $f_{\rm exc}(\varepsilon)=(1/2)\int_{-1}^{1}du\,f(\sqrt{2m\varepsilon},u)$, where $u=\cos\theta$. These non-analyticities are always present for the different types of heat baths considered here, but their degree and location depend on the bath type. For the gapped bath, we have found that the non-analytic surfaces are located at momenta given by (see Fig.~\ref{fig2}(a)):
\begin{align}
&{\bf p}^2_i/2m=\varepsilon_i=l \Omega+s\Delta.
\label{surfaces-gapped}
\end{align}
where $\Omega$ is the driving frequency, and $i$ is a short-hand for two indices $i\leftrightarrow (l,s)$ that label the surfaces. $(l,s)$ can be any integers as long as the right-hand side is positive. We have found that the strongest non-analyticities tend to occur for $(l,s)\in \{(0,1),(1,-1),(1,1)\}$, and, thus, we will focus on these \footnote{For a discussion of other weaker non-analyticities see SM Sec.~\ref{S-surface-hierarchy}.}. Amazingly, we find that in the non-equilibrium BEC state these strongest non-analyticities correspond to actual jumps of $f_{\bf p}$ that remain sharp even at finite temperature (see Fig.~\ref{fig2}(a)). In the non-equilibrium normal state, these become square-root kinks of the derivative of $f_{\bf p}$ (see Fig.~\ref{fig2}(b)), which is the same type of non-analyticity observed for Fermions~\cite{shi2025ultracritical}. Further details of these jumps are discussed in SM Sec.~\ref{S-gapped-bath-author-notation}. 

\noindent For the ohmic bath we find non-analyticities instead at:
\begin{align}
&{\bf p}^2_i/2m=\varepsilon_i=l \Omega.
\label{surfaces-ohmic}
\end{align}
where $l$ is a positive integer. In this case, these are jumps of the derivative of the occupation in the BEC state (see Fig.~\ref{fig2}(c)), and in the normal state become square-root kinks of the second derivative of the occupation. The different behaviors are summarized in the Table in Fig.~\ref{fig2}(d), and for more details see SM Secs.~\ref{S-ohmic-nonanalyticities}--\ref{S-surface-hierarchy}. 

\smallskip

\begin{figure}[!htbp]
\centering
\includegraphics[width=\columnwidth]{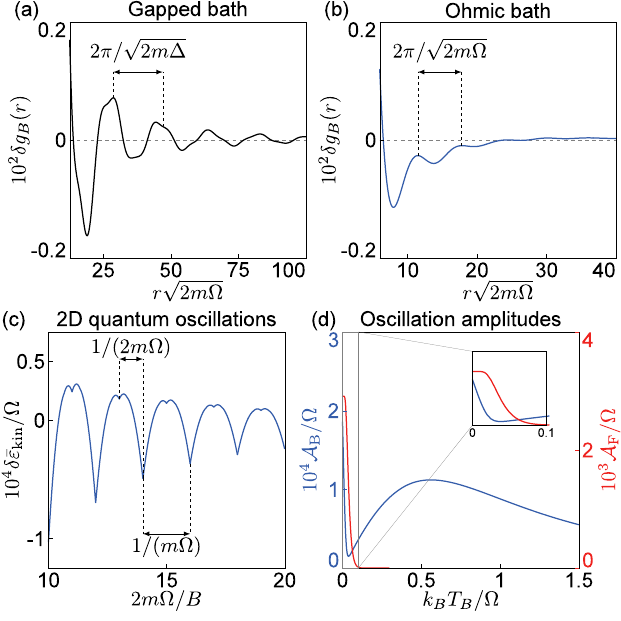}
\caption{(a,b) Background-subtracted density correlations $\delta g_B(r)=C_B(r)/n_{\rm tot}^2-\hat C_g/r$ of a 3D driven BEC for (a) gapped and (b) ohmic baths. The coefficient $\hat C_g$ is fitted to the long-distance correlation (SM Sec.~\ref{S-C-driven-background}). The measurement direction is at $\arccos(1/\sqrt3)\simeq54.7^\circ$ to the normal to the drive plane. Arrows indicate the characteristic surface wavelengths.
(c) Oscillatory part of time-averaged kinetic energy per particle in 2D driven normal Bose gas coupled to ohmic bath.
(d) Temperature dependence of amplitude of oscillations in (c) (blue, left scale) and an undriven equilibrium Fermi gas at the same density and temperature $T=T_B$ (red, right scale).
Parameters in (a,b) are as in Fig.~\ref{fig2}(a,c), respectively, except $A_0/\sqrt{2m\Omega}=0.5$, $k_BT_B/\Omega=0.005$, and $n_{\rm tot}/(2m\Omega)^{3/2}=0.03$. The Bose parameters in (c,d) are as in Fig.~\ref{fig2}(c), except $n_{\rm tot}/(2m\Omega)=0.2/\pi$, with the common field interval $10\le 2m\Omega/B\le20$; in (c), $k_BT_B/\Omega=0.2$.}
\label{fig3}
\end{figure}

\noindent \textit{\color{blue}Long-ranged oscillatory density correlations.} We will now substantiate the idea that these non-analyticities behave like emergent Fermi surfaces by showing that they lead to analogous observable fingerprints. A characteristic fingerprint of Fermi surfaces are the oscillatory power-law decaying spatial correlations of observables with a period controlled by their Fermi radius \cite{giuliani2005quantum}. To investigate these, we compute the equal-time two-point boson-density correlation function: $C_B({\bf r})=\langle\hat n({\bf0})\hat n({\bf r})\rangle-\langle\hat n({\bf0})\rangle\langle\hat n({\bf r})\rangle$. In SM Sec.~\ref{S-C-bunching}, we show that this correlator is related to the Fourier transform of the occupation of excited states, $\tilde f_{\rm exc}({\bf r})=V^{-1}\sum_{\bf p\neq0} e^{i{\bf p}\cdot{\bf r}}f_{\bf p}$ \footnote{Which is a real function because the occupation is inversion symmetric $f_{\bf p}=f_{\bf -p}$.}, and the condensate density, $n_0=V^{-1}f_{\bf p=0}$, as follows (${\bf r}\ne0$):
\begin{align}
&C_B({\bf r})=2n_0\tilde f_{\mathrm{exc}}({\bf r})+\tilde f_{\mathrm{exc}}^2({\bf r}).
\label{eq:pair-correlation}
\end{align}
Therefore, the non-analyticities of $f_{\bf p}$ at finite momenta are expected to lead to long-ranged oscillating  correlations both in the BEC ($n_0\neq 0$) and normal states ($n_0 = 0$), albeit with different power laws and periods. For example, for the BEC state coupled to a gapped bath, by writing the occupation jumps as $\sum_i A_i(\hat{\bf p})\Theta(p_{F,i}-p)$, one gets that the asymptotic oscillatory part of the correlations is:
\begin{align}
&[C_B({\bf r})]_{\rm osc}\simeq
-\frac{n_0}{\pi^2 r^2}
\sum_i A_i(\hat{\bf r})p_{F,i}\cos(p_{F,i}r).
\label{eq:density-oscillations}
\end{align}
Fig.~\ref{fig3}(a,b), which is obtained from the full numerical calculation of Eq.\eqref{eq:pair-correlation} after subtracting a fitted leading $1/r$ background, illustrates this behavior. We have also obtained analogous oscillations in the case of the BEC coupled to the gapless ohmic bath but with a $1/r^3$  power-law decay. Notice that these oscillations have a slower power-law decay than even the ideal three-dimensional Fermi gas at zero temperature, which decay as $1/r^4$ and instead oscillate with $2p_F$ wavevector~\cite{giuliani2005quantum}. In the non-equilibrium normal states of bosons, the oscillations would also have wavevector $2p_{F,i}$, and their power law of decay would faster and coincide with that of Fermions \cite{shi2025ultracritical}. Further details of the oscillations are discussed in SM Sec.~\ref{S-C-driven-background}.

\smallskip

\noindent \textit{\color{blue}Quantum Oscillations.} To demonstrate that these emergent Fermi surfaces also give rise to quantum oscillations in response to external magnetic fields, we will focus on the two-dimensional normal state \footnote{Because the effects of Landau quantization are stronger in 2D compared to 3D. Moreover, being in the normal state is needed to avoid the formation of the Abrikosov lattice formation.}. We take the magnetic field ${\bf B}=-B\hat{\bf z}$ ($B>0$) to be oriented perpendicular to the system $xy$-plane. We use the exact single-particle Landau-level Floquet spectrum  ~\cite{shi2025algebraic} and re-derive and solve numerically the Floquet-Boltzmann equation for the steady-state occupations of Landau levels, analogously to Ref.~\cite{matsyshyn2025high}, as detailed in SM Sec.~\ref{S-2DLandau}. Figure~\ref{fig3}(c) shows the time-averaged kinetic-energy per particle as a function of $1/B$ for the case of the ohmic bath, after subtracting the  non-oscillating background (see SM Sec.~\ref{S-fig3-background} for details). The main dips have the expected period $1/(m\Omega)$ from the Fermi surface at $\varepsilon=\Omega$. The intervening smaller dents with period $1/(2m\Omega)$ arise from the surface at $\varepsilon=2\Omega$. Remarkably, as shown in Fig.~\ref{fig3}(d), the quantum oscillations survive up to extremely high temperatures comparable to the frequency of the drive,  in analogy to what was found for the case of fermions \cite{matsyshyn2025high}. This contrasts sharply with the standard Lifshitz-Kosevich behavior of the equilibrium quantum oscillations of Fermions \cite{shoenberg1984magnetic} which are exponentially suppressed for temperatures above the cyclotron energy, and are shown in red in Fig.~\ref{fig3}(d) for comparison. 

\smallskip

\noindent \textit{\color{blue}Ultra-cold atomic realization.} 
Ultra-cold atoms offer promising platforms to experimentally discover these non-equilibrium fermi surfaces of Bose gases. A potential platform could consist of two atomic clouds: a larger one acting as a bath, and a smaller one as the bosonic system of interest. The non-equilibrium driving of this system could be performed by rotating its trapping center at frequency $\Omega$, using specie selective trapping and driving strategies \cite{LeBlanc2007,Catani2012,Lamporesi2010,scelle2013motional,rentrop2016observation,catani2009entropy,mckay2013slow,chen2014bath}.  Small driving frequencies are desirable as long as they remain larger than inter-atomic scattering rates \footnote{Because the strength of Floquet umklapp processes increases with the dimensionless amplitude of driving $F_0 v/(\hbar \Omega^2)$, where $F_0$ is the amplitude of the driving force and $v$ is the typical velocity of atoms in the trap. But the frequency should remain larger than the scattering rates to avoid quasi-particle lifetime effects smearing the sharpness of these Fermi surfaces}. Given that typical inter-atomic collision times are on the order of $\sim 1$ ms~\cite{catani2009entropy}, driving at $f = \Omega/2\pi \sim 10$ kHz should meet this requirement.

The Friedel-like oscillating ripples of the density correlations could be used as a fingerprint to detect these states. Their wavelength is expected to double upon cooling from the non-equilibrium normal state into the non-equilibrium BEC:   $\lambda_{\rm BEC} =2 \lambda_{\rm normal} \sim \sqrt{\pi h/(m \Omega)}$  \footnote{These is a rough estimate assuming that our ohmic gappless bath of harmonic oscillators, can be used as a proxy for the more complex real bath created by the second larger atomic cloud. The amplitude and visibility oscillations should become smaller in the normal state compared to the BEC.}. Notice that these wave-lengths are independent of the boson density, and, therefore, its spatial variation across the cloud should not be particularly detrimental to their coherence \footnote{This is a great advantage for observing these oscillations in non-uniform cold-atomic clouds in comparison to standard equilibrium Fermionic Friedel oscillations \cite{riechers2017detecting}, whose period changes with density.}. For  $^{39}\mathrm{K}$ at driving frequency $f=\Omega/2\pi \sim 10$ kHz, this length is about $\lambda_{\rm BEC} \sim 0.7 \mu$m. Such oscillatory density correlations could be detected using existing imaging techniques \cite{Xiang2025,Yao2025,Jongh2025,Hertkorn2021,hartke2023direct,tamura2026direct,folling2005spatial,jeltes2007comparison,manz2010two,fang2016momentum,jacqmin2011sub}. Alternatively, external potentials could be used to induce standing-wave ripples, such as by having an extra atomic impurity or polaron \cite{Jorgensen2016,Hu2016,Schmidt2018,etrych2025universal}, additional weak sinusoidal optical potential \cite{Tao2023,chauveau2023superfluid}, focused optical spot \cite{Andrews1997,Raman,Onofrio2000,gauthier2016direct}, optical planar barriers \cite{potnis2017interaction,Albiez2005,levy2007ac}, or creating box-like trapping  boundaries for the cloud  \cite{gaunt2013bose,Meyrath2005,chomaz2015emergence}. 

Another strategy is to detect quantum oscillations created by effective magnetic fields obtained from rotating quasi-2D clouds \cite{schweikhard2004rapidly,fletcher2021geometric,sharma2024thermal,abo2001observation} or other strategies \cite{leblanc2012observation,lin2009synthetic,kennedy2015observation,aidelsburger2015measuring,Aidelsburger}. As we have shown, the non-equilibrium quantum oscillations can survive up to extremely high temperatures comparable to the driving frequency $\Omega$ (see Fig.~\ref{fig3}(d)), which can be chosen to exceed their BKT temperature \cite{Prokofev2001}. Fig.~\ref{fig3}(d), illustrates that the amplitude of non-equilibrium  oscillations of kinetic energy can be even just one order of magnitude smaller than the analogous equilibrium oscillations of a Fermi gas. 

Finally, since the emergent Fermi surfaces in the driven normal state of bosons are expected to become stronger in lower dimensions~\cite{shi2025ultracritical}, quasi-1D bosonic clouds are also promising platforms. The recent report of Friedel-like oscillatory correlations in a non-equilibrium 1D bosonics ``fractional Fermi sea''~\cite{zeng2026realization}, makes us optimistic that analogous detection of Friedel-like oscillations of our proposed non-equilibrium Fermi surfaces is also  within reach. 
 
All of the above highlights that there is a wealth of technologies and strategies currently available in ultra-cold atomic platforms for discovering and investigating these exotic non-equilibrium states of bosons.

\textit{Acknowledgements.} We acknowledge support by the Deutsche Forschungsgemeinschaft (DFG) through Research Grant Projects No. 542614019, No. 518372354, and No. 555335098.

\nocite{riechers2017detecting}
\bibliography{ref}

\clearpage

\setcounter{secnumdepth}{3}
\appendix

\renewcommand{\theequation}{\thesection-\arabic{equation}}
\renewcommand{\thefigure}{\thesection-\arabic{figure}}
\renewcommand{\theHfigure}{\thesection.\arabic{figure}}
\renewcommand{\thetable}{\thesection-\Roman{table}}

\onecolumngrid

\section*{Supplementary Material for ``Emergent Fermi Surfaces in Periodically Driven Bose Gases''}

\section{Floquet-Boltzmann Equation for a Driven Bose Gas with Condensation}
\label{S-kinetics}
\setcounter{figure}{0}

This section gives the derivation of the Floquet-Boltzmann equation for a driven {\it bosonic} system coupled to a {\it bosonic bath}, with a special emphasis on the treatment of a Bose-Einstein condensate. While the general formalism of the quantum kinetic equation shares similarities with the fermionic case, we highlight the crucial differences arising from bosonic statistics and the presence of a macroscopic condensate.

\subsection{From the Exact Liouville Dynamics to the Floquet-Fermi Golden Rule}
\label{FFGR}

To establish the non-equilibrium kinetic framework for the driven bosonic system, we derive the collision integral directly from the microscopic dynamics.
We write the total Hamiltonian as $\hat H(t)=\hat H_0(t)+\hat V(t)$, with $\hat H_0(t)=\hat H_S(t)+\hat H_B$ and $\hat V(t)\equiv\hat H_{SB}$ the system-bath coupling introduced in the main text. In the bath-mode notation used here, $\hat H_B=\sum_q\omega_q\hat b_q^\dagger\hat b_q$; for the independent local baths, the mode sum includes the spatial integral. In an arbitrary one-particle basis for the system, the density coupling reads:
\begin{align}
\begin{aligned}
&\hat{V}(t) = \sum_{q,\nu, \eta} \bra{\nu} \hat{\chi}_{q} \ket{\eta} \hat{b}_q \hat{a}_\nu^\dagger \hat{a}_\eta + \text{h.c.} = \sum_q \big( \hat{\mathcal{V}}_q \hat{b}_q + \hat{\mathcal{V}}_q^\dagger \hat{b}_q^\dagger \big).
\end{aligned}
\end{align}
Here, the italic index $q$ labels a bath mode and is distinct from the system momentum ${\bf q}$. The operator $\hat{\mathcal{V}}_q \equiv \sum_{\nu, \eta} \bra{\nu} \hat{\chi}_{q} \ket{\eta} \hat{a}_\nu^\dagger \hat{a}_\eta$ is the system scattering operator coupled to bath mode $q$, and $\hat{\chi}_{q}$ defines its corresponding interaction vertex acting in the single-particle Hilbert space. We write the quadratic system Hamiltonian as $\hat H_S(t)=\sum_{\nu\eta}h_{\nu\eta}(t)\hat a_\nu^\dagger\hat a_\eta$, distinguishing its one-particle operator $\hat h(t)$ from the Fock-space operator $\hat H_S(t)$. Collisions between gas particles are omitted, as in the main-text kinetic model; for a weakly interacting realization their collision rate must be much smaller than the bath-induced relaxation rate.

The system operator in every term is the number-conserving bilinear
$\hat a_\nu^\dagger\hat a_\eta$. The coupling can therefore exchange energy
and momentum with the bath without changing the total number of driven
bosons.
 The joint system--bath density matrix $\hat{\rho} (t)$ obeys the exact Liouville equation:
\begin{align}
\begin{aligned}
&i d_t\hat{\rho} (t) = [\hat{H}_0(t), \hat{\rho} (t)] + [\hat{V}(t), \hat{\rho} (t)].
\end{aligned}
\end{align}
In the interaction picture, $\partial_t\rho_I(t)=-i[V_I(t),\rho_I(t)]$. Integrating once gives $\rho_I(t)=\rho_I(t_3)-i\int_{t_3}^{t}dt'\,[V_I(t'),\rho_I(t')]$. Substituting this identity into the differential equation and returning to the Schr\"odinger picture gives the exact kinetic equation:
\begin{align}
\begin{aligned}
&d_t\hat{\rho} (t) = -i[\hat{H}_0(t),\hat{\rho} (t)] -i\big[\hat{V}(t), U_0(t,t_3)\hat{\rho}(t_3)U_0(t_3,t)\big] \\
&\quad + \left(-i\right)^2\int_{t_3}^t d t^{\prime}\Big[\hat{V}(t),\big[U_0(t,t^{\prime})\hat{V}(t^{\prime})U_0(t^{\prime},t), U_0(t,t^{\prime})\hat{\rho}(t^{\prime})U_0(t^{\prime},t)\big]\Big].
\end{aligned}
\end{align}
Here, $U_0(t,t^{\prime}) = \mathcal{U}_S(t,t^{\prime}) \otimes \mathcal{U}_B(t,t^{\prime})$ is the unitary evolution operator acting on the joint system--bath Fock space. Because the interaction picture is defined with respect to the non-interacting Hamiltonian, the system and bath commute ($[\hat{H}_S(t), \hat{H}_B] = 0$), allowing the evolution operator to factor into this uncoupled tensor product. At the initial time $t_3 \to -\infty$, we assume an initial decorrelation $\hat{\rho}(t_3) = \hat{\rho}_S(t_3) \otimes \hat{\rho}_B$, where $\hat{\rho}_B$ is the stationary thermal equilibrium state of the bath. Tracing over the bath degrees of freedom ($\text{Tr}_B$), the term linear in $\hat{V}(t)$ vanishes since $\langle \hat{b}_q \rangle_B = \langle \hat{b}_q^\dagger \rangle_B = 0$.

The trace over the nested double commutator requires the evaluation of the thermal bath correlation functions. Using the free evolution of the bath annihilation operators $\mathcal{U}_B^{\dagger}(t,t^{\prime}) \hat{b}_q \mathcal{U}_B(t,t^{\prime}) = e^{-i\omega_q(t-t^{\prime})} \hat{b}_q$, the expectations yield $\langle \hat{b}_q \hat{b}_{q^{\prime}}^{\dagger} \rangle_B = \delta_{qq^{\prime}} \big(1+N_b(\omega_q)\big)$ for emission processes, and $\langle \hat{b}_q^{\dagger} \hat{b}_{q^{\prime}} \rangle_B = \delta_{qq^{\prime}} N_b(\omega_q)$ for absorption processes.

To obtain the kinetic equation for the single-particle density matrix $\hat{n}(t)$, whose matrix elements are defined as $n_{\gamma \delta}(t) \equiv \text{Tr}_{\text{Fock}} [\hat{\rho}(t) \hat{a}_\delta^{\dagger} \hat{a}_\gamma]$, we evaluate the trace over the joint system--bath Fock space. This yields an equation for one-particle observables in the single-particle Hilbert space. 

The connection between these spaces is established via the Heisenberg evolution of the annihilation operator under the free system Hamiltonian: $\mathcal{U}_S^\dagger(t,t') \hat{a}_\alpha \mathcal{U}_S(t,t') = \sum_\beta \big[U_S(t,t')\big]_{\alpha\beta} \hat{a}_\beta$, where $U_S(t,t')$ is now the single-particle unitary evolution matrix. At second order in the system--bath coupling, we use the Born
factorization $\hat\rho(t')\simeq\hat\rho_S(t')\otimes\hat\rho_B$. Let
$\hat O_{\gamma\delta}=\hat a_\delta^\dagger\hat a_\gamma$. Writing the two
vertices in the interaction picture, cyclicity of the full Fock-space trace
gives the operator averages to be evaluated:
\begin{align}
\begin{aligned}
&{\rm Tr}_{S,B}\!\left\{ \hat O_{\gamma\delta} [\hat V(t),[\hat V(t'),\hat\rho_S(t')\otimes\hat\rho_B]] \right\} =\left\langle [[\hat O_{\gamma\delta},\hat V(t)],\hat V(t')] \right\rangle.
\end{aligned}
\end{align}
The commutator of the measured one-body operator with a system vertex remains
a number-conserving bilinear, since
\begin{align}
\begin{aligned}
&[\hat a_\delta^\dagger\hat a_\gamma, \hat a_\nu^\dagger\hat a_\eta] =\delta_{\gamma\nu}\hat a_\delta^\dagger\hat a_\eta -\delta_{\eta\delta}\hat a_\nu^\dagger\hat a_\gamma.
\end{aligned}
\end{align}
After the bath contractions in the preceding paragraph, representative system
terms therefore have the form
$\langle[\hat O_{\gamma\delta},\hat{\mathcal V}_q(t)]
\hat{\mathcal V}_q^\dagger(t')\rangle_S$, together with the reversed ordering
and Hermitian-conjugate terms. Each of these contains four system creation or annihilation
operators. In contrast, the uncommuted product
$\langle\hat a_\delta^\dagger\hat a_\gamma
\hat{\mathcal V}_q(t)\hat{\mathcal V}_q^\dagger(t')\rangle_S$ contains six
system creation or annihilation operators and is not the four-operator average evaluated below.

We describe the excited particles as free bosons with independent mode occupations, neglecting correlations between distinct modes. Four-operator averages are evaluated as sums of products of pair averages, with $\langle\hat a_i\rangle=\langle\hat a_i\hat a_j\rangle=0$. The same free-boson description is used for the density correlations in SM Sec.~\ref{S-C-bunching}. For arbitrary one-particle matrices $A$ and $B$, this gives
\begin{align}
\begin{aligned}
&\left\langle(\hat a^\dagger A\hat a)(\hat a^\dagger B\hat a)\right\rangle ={\rm Tr}_{\rm sp}(A n){\rm Tr}_{\rm sp}(B n) +{\rm Tr}_{\rm sp}[A(1+n)B n],
\end{aligned}
\end{align}
in which the subscript ``${\rm sp}$'' stands for single particle.
It follows by inserting $\langle\hat a_i^\dagger\hat a_j\hat a_k^\dagger\hat a_l\rangle=n_{ji}n_{lk}+n_{li}(\delta_{jk}+n_{jk})$. Combining the operator orderings
fixed by the nested commutator and normal-ordering with
$[\hat a_j,\hat a_k^\dagger]=\delta_{jk}$ gives
$\langle\hat a_j\hat a_k^\dagger\rangle=\delta_{jk}+n_{jk}$. For the
translationally invariant local-density coupling used below, the direct
contraction drops out after summing the identical local baths: for nonzero
momentum transfer its trace vanishes, whereas the zero-momentum vertex is
proportional to the identity and vanishes inside the commutator. The exchange contraction then gives the bosonic factors $\hat n(t')$ and $1+\hat n(t')$. For a condensate, only the factorization of occupations of distinct modes is needed: $\langle N_{\bf p}N_{\bf q}\rangle\simeq f_{\bf p}f_{\bf q}$ for ${\bf p}\ne{\bf q}$. The diagonal transition cancels from the population equation, so no thermal value of the condensate number variance is imposed.

Evaluating these free-boson averages reduces the Fock-space expectation to a trace over the single-particle basis (denoted as ${\rm Tr}_{\rm sp}$). Tracking the terms proportional to $(1+N_b(\omega_q))$ and applying the cyclic property of ${\rm Tr}_{\rm sp}$ to isolate the external state projector $\ket{\delta}\bra{\gamma}$, we reconstruct the bosonic formulation of the emission integral~\cite{shi2025ultracritical}. In particular, with $O=|\delta\rangle\langle\gamma|$, $\tilde n=U_S n(t')U_S^\dagger$, and $Y=U_S\chi_q^\dagger U_S^\dagger$, cyclicity gives
\begin{align}
\begin{aligned}
&-{\rm Tr}_{\rm sp}\{[O,\chi_q](1+\tilde n)Y\tilde n\} ={\rm Tr}_{\rm sp}\{O[(1+\tilde n)Y\tilde n,\chi_q]\}.
\end{aligned}
\end{align}
The other emission ordering is its Hermitian conjugate. Since $(1+\tilde n)Y\tilde n=U_S(1+n(t'))\chi_q^\dagger n(t')U_S^\dagger$, this yields
\begin{align}
\begin{aligned}
&\hat{I}^{\text{e}}(t) = \sum_q \int_{-\infty}^t d t^{\prime} e^{i\omega_q(t^{\prime}-t)} \big(1+N_b(\omega_q)\big) \Big[ U_S(t,t^{\prime})\big(1+\hat{n}(t^{\prime})\big)\hat{\chi}_{q}^{\dagger} \hat{n}(t^{\prime}) U_S^{\dagger}(t,t^{\prime}), \hat{\chi}_{q} \Big] + \text{h.c.}
\end{aligned}
\end{align}
Here, $U_S(t,t^{\prime})$, $\hat{n}(t)$, and $\hat{\chi}_{q}$ act in the single-particle Hilbert space.

We now evaluate the diagonal component of this collision operator in the time-dependent Floquet basis. These basis states are defined by the free single-particle Schrödinger equation, $i \partial_t |\psi_{\mathbf{p}}(t)\rangle = \hat h(t) |\psi_{\mathbf{p}}(t)\rangle$, ensuring that a diagonal density matrix constructed from them represents the solution to the collisionless dynamics, $i d_t\hat{n} (t) = [\hat h(t), \hat{n} (t)]$. The driven single-particle Hamiltonian possesses Floquet states $|\psi_{\bf p}(t)\rangle = e^{-i\varepsilon_{\bf p} t} |\phi_{\bf p}(t)\rangle$, where $\varepsilon_{\bf p}$ is the quasienergy. We use the population approximation in the Floquet basis, discarding coherences that average out on the kinetic time scale. In the translationally invariant state considered here, the one-body density matrix is diagonal in momentum. The Markov step holds the slowly varying occupations fixed during the collision integral, $f_{\bf p}(t')\simeq f_{\bf p}(t)$, while retaining the free evolution of the basis states. In a periodic steady state,
\begin{align}
\begin{aligned}
&\hat{n}(t) = \sum_{\bf p} f_{\bf p} |\psi_{\bf p}(t)\rangle \langle \psi_{\bf p}(t)|.
\end{aligned}
\end{align}
In this basis, the single-particle evolution operator is $U_S(t,t^{\prime}) = \sum_{\bf p} |\psi_{\bf p}(t)\rangle \langle \psi_{\bf p}(t^{\prime})|$. Evaluating the diagonal matrix elements $\langle \psi_{\bf p}(t) | \hat{I}^{\text{e}}(t) | \psi_{\bf p}(t) \rangle$, the internal propagation operators $U_S(t,t^{\prime})$ simplify the expression by shifting the external states to time $t^{\prime}$. We use the scattering rates $W$ of the main text throughout. To keep track of normalization in this derivation, we temporarily normalize the Floquet states in a box of volume $V$: $W_{{\bf q}\to{\bf p}}/V$ is the transition probability per unit time between two box-normalized states. The instantaneous emission contribution in this convention is
\begin{align}
\begin{aligned}
&W^{\text{e}}_{{\bf q} \to {\bf p}}(t) = V\sum_q \big(1+N_b(\omega_q)\big) \int_{-\infty}^t d t^{\prime} e^{i\omega_q(t^{\prime}-t)} \langle \psi_{\bf p}(t^{\prime}) | \hat{\chi}_{q}^{\dagger} | \psi_{\bf q}(t^{\prime}) \rangle \langle \psi_{\bf q}(t) | \hat{\chi}_{q} | \psi_{\bf p}(t) \rangle + \text{h.c.}
\end{aligned}
\end{align}
Using Floquet's theorem, we separate the secular quasienergy phases from the periodic micromotion:
\begin{align}
\begin{aligned}
&\langle \psi_{\bf q}(t) | \hat{\chi}_q | \psi_{\bf p}(t) \rangle = e^{i(\varepsilon_{\bf q} - \varepsilon_{\bf p})t} \chi^q_{{\bf q}, {\bf p}}(t) = e^{i(\varepsilon_{\bf q} - \varepsilon_{\bf p})t} \sum_{l=-\infty}^{\infty} \chi^q_{{\bf q}, {\bf p}}(l) e^{-il\Omega t},
\end{aligned}
\end{align}
where $\chi^q_{{\bf q}, {\bf p}}(l) = \sum_{l^{\prime}} \langle \varphi_{l^{\prime}, {\bf q}} | \hat{\chi}_q | \varphi_{l^{\prime}+l, {\bf p}} \rangle$ represents the Floquet-Fourier harmonic of the single-particle interaction vertex. Inserting this expansion into the emission rate and applying the change of variable $\tau = t^{\prime} - t$, the remaining time integral is independent of the external time $t$:
\begin{align}
\begin{aligned}
&W^{\text{e}}_{{\bf q} \to {\bf p}}(t) = V\sum_q \sum_{l, l^{\prime}} \big(1+N_b(\omega_q)\big) e^{i(l^{\prime}-l)\Omega t} \chi^q_{{\bf q}, {\bf p}}(l) \big[\chi^q_{{\bf q}, {\bf p}}(l^{\prime})\big]^* \int_{-\infty}^0 d\tau e^{i\big(\omega_q + (\varepsilon_{\bf p} - \varepsilon_{\bf q}) + l^{\prime}\Omega\big)\tau} + \text{h.c.}
\end{aligned}
\end{align}
In the physically relevant weak-coupling limit, the macroscopic populations $f_{\bf p}$ evolve on timescales vastly exceeding the driving period $T = 2\pi/\Omega$. Thus, the macroscopic non-equilibrium steady state is governed by the temporal average of the collision integral over one driving period. This coarse-graining uses $T^{-1}\int_0^Tdt\,e^{i(l'-l)\Omega t}=\delta_{ll'}$ to remove interference between different drive harmonics. The subsequent time integral uses
\begin{align}
\begin{aligned}
&\lim_{\eta\to0^+}\int_{-\infty}^{0}d\tau\,e^{\eta\tau+iE\tau} =\pi\delta(E)-i\,P\frac{1}{E}.
\end{aligned}
\end{align}
Adding the Hermitian conjugate gives $2\pi\delta(E)$ in the population rate. The principal-value term represents a second-order energy shift, omitted from the leading kinetic description. The resulting Floquet-Fermi golden rule is
\begin{align}
\begin{aligned}
&\langle W^{\text{e}}_{{\bf q} \to {\bf p}} \rangle_T = V\sum_q \big(1+N_b(\omega_q)\big) \sum_{l=-\infty}^{\infty} |\chi^q_{{\bf q}, {\bf p}}(l)|^2 \int_{-\infty}^0 d\tau e^{i\big(\omega_q + (\varepsilon_{\bf p} - \varepsilon_{\bf q}) + l\Omega\big)\tau} + \text{h.c.} \\
&\quad = V\sum_q {2\pi} \big(1+N_b(\omega_q)\big) \sum_{l=-\infty}^{\infty} |\chi^q_{{\bf q}, {\bf p}}(l)|^2 \delta\big(\omega_q + \varepsilon_{\bf p} - \varepsilon_{\bf q} + l\Omega\big).
\end{aligned}
\end{align}
The corresponding cycle-averaged absorption contribution to the scattering rate $\langle W^{\text{a}}_{{\bf q} \to {\bf p}} \rangle_T$ is obtained identically by the substitutions $1+N_b(\omega_q) \to N_b(\omega_q)$, $\omega_q \to -\omega_q$, and $\hat{\chi}_q \leftrightarrow \hat{\chi}_q^\dagger$.

To combine emission and absorption over the continuous bath spectrum, we define the density of modes of one local continuum as $\nu_B(\omega)=\sum_\eta\delta(\omega-\omega_\eta)$, with $\eta$ labeling its internal modes, and use $S(\omega)=[N_b(|\omega|)+\Theta(\omega)]\nu_B(|\omega|)$. Positive $\omega$ denotes energy emitted into the bath. The Bose factors from the density-matrix calculation give the finite-volume Floquet-Boltzmann equation
\begin{align}
\begin{aligned}
&\partial_t f_{\bf p}=\frac1V\sum_{\bf q}\left[f_{\bf q}W_{{\bf q}\to{\bf p}}(1+f_{\bf p})-f_{\bf p}W_{{\bf p}\to{\bf q}}(1+f_{\bf q})\right],
\end{aligned}
\end{align}
where $W=\langle W^{\rm e}+W^{\rm a}\rangle_T$. The steady-state equation is therefore
\begin{align}
\begin{aligned}
&0 = \frac1V\sum_{\bf q} \Big[ f_{\bf q} W_{{\bf q} \to {\bf p}} (1 + f_{\bf p}) - f_{\bf p} W_{{\bf p} \to {\bf q}} (1 + f_{\bf q}) \Big].
\end{aligned}
\end{align}
For our system-bath configuration, an independent continuum couples to the local particle density at each point, with vertex $\hat\chi_{({\bf r},\eta)}=\chi_0|{\bf r}\rangle\langle{\bf r}|$. Independence means $[b_\eta({\bf r}),b_{\eta'}^\dagger({\bf r}')]=\delta_{\eta\eta'}\delta^{(3)}({\bf r}-{\bf r}')$, so rates from different points add. For $\langle{\bf r}|\phi_{\bf p}(t)\rangle=V^{-1/2}e^{i{\bf p}\cdot{\bf r}}\phi_{\bf p}(t)$, define
\begingroup
\begin{align}
\begin{aligned}
&u_{{\bf p}{\bf q}}^{(l)}=\frac1T\int_0^Tdt\,e^{il\Omega t}\phi_{\bf p}^*(t)\phi_{\bf q}(t) =\sum_a\varphi_{a,{\bf p}}^*\varphi_{a+l,{\bf q}},\qquad \int_Vd^3r\left|\frac{\chi_0}{V}e^{i({\bf q}-{\bf p})\cdot{\bf r}}u_{{\bf p}{\bf q}}^{(l)}\right|^2 =\frac{|\chi_0|^2}{V}\Phi_{{\bf q},{\bf p}}^{(l)}, \\
&\Phi_{{\bf q},{\bf p}}^{(l)}=|u_{{\bf p}{\bf q}}^{(l)}|^2 =\Big|\sum_a\varphi_{a,{\bf p}}\varphi_{a+l,{\bf q}}^*\Big|^2.
\end{aligned}
\label{S-Floquet-overlap}
\end{align}
\endgroup
\begingroup
For the plane-wave modes considered here, $|\varphi_{a,{\bf p}}\rangle=\varphi_{a,{\bf p}}|{\bf p}\rangle$, with $\langle{\bf r}|{\bf p}\rangle=V^{-1/2}e^{i{\bf p}\cdot{\bf r}}$ and $\langle{\bf p}|{\bf p}\rangle=1$. Equivalent product and trace forms are
\begin{align}
\Phi^{(l)}_{{\bf q},{\bf p}}
&={\rm Re}\sum_{a,b}
\big(\varphi_{a,{\bf p}}\varphi_{b,{\bf p}}^*\big)
\big(\varphi_{b+l,{\bf q}}\varphi_{a+l,{\bf q}}^*\big)
\label{S-Floquet-weight-product}\\
&={\rm Tr} \Big\{
\Big[\sum_a|\varphi_{a,{\bf p}}\rangle\langle\varphi_{a+l,{\bf q}}|\Big]
\Big[\sum_b|\varphi_{b,{\bf p}}\rangle\langle\varphi_{b+l,{\bf q}}|\Big]^\dagger
\Big\}.
\label{S-Floquet-weight-trace}
\end{align}
For the four-harmonic product $\varphi_{a,{\bf p}}\varphi_{b,{\bf p}}^*\varphi_{c,{\bf q}}\varphi_{d,{\bf q}}^*$, cycle averaging imposes $a+c=b+d$. Setting $l=c-b$ fixes $c=b+l$ and $d=a+l$, leaving ${\rm Re}([u_{{\bf p}{\bf q}}^{(l)}]^*u_{{\bf p}{\bf q}}^{(l)})=|u_{{\bf p}{\bf q}}^{(l)}|^2$. The second line has the trace-of-product structure of Eq.~(8) of Ref.~\cite{matsyshyn2025high}; its equality to $\Phi^{(l)}_{{\bf q},{\bf p}}$ here follows directly from $\sum_a|\varphi_{a,{\bf p}}\rangle\langle\varphi_{a+l,{\bf q}}|=[u_{{\bf p}{\bf q}}^{(l)}]^*|{\bf p}\rangle\langle{\bf q}|$.
More generally, without assuming plane-wave factorization, define the harmonic blocks of the Floquet projector
$\hat P_{\bf p}^{ab}=|\varphi_{a,{\bf p}}\rangle\langle\varphi_{b,{\bf p}}|$. The same identity for an arbitrary interaction vertex reads
\begin{align}
\left|\chi^q_{{\bf q},{\bf p}}(l)\right|^2
&=\Big|\sum_a\langle\varphi_{a,{\bf q}}|\hat\chi_q|\varphi_{a+l,{\bf p}}\rangle\Big|^2
=\sum_{a,b}{\rm Tr} \Big[
\hat P_{\bf q}^{ba}\hat\chi_q\hat P_{\bf p}^{a+l,b+l}\hat\chi_q^\dagger
\Big],
\label{S-Floquet-vertex-projectors}
\end{align}
where the trace is over the single-particle Hilbert space. This is the coherent vertex factor of Ref.~\cite{matsyshyn2025high}, written as a product of projector blocks, and holds for arbitrary periodic driving and interaction form factor. The identities retain the off-diagonal harmonic terms $a\ne b$: cycle averaging equates the net sideband indices but does not impose $a=b$. For the circular drive and local bath considered here, the relative-azimuthal average additionally removes these terms, yielding $\overline\Phi^{(l)}_{{\bf q},{\bf p}}=\sum_a|\varphi_{a,{\bf p}}|^2|\varphi_{a+l,{\bf q}}|^2$, as shown in Eq.~\eqref{angle-averaged-rates}.
\par\endgroup
Here $\phi_{\bf p}(t)=\sum_a\varphi_{a,{\bf p}}e^{-ia\Omega t}$. In the emission rate above, relabel $l\to-l$ and use $[u_{{\bf q}{\bf p}}^{(-l)}]^*=u_{{\bf p}{\bf q}}^{(l)}$; its delta function then selects $\omega_q=\varepsilon_{\bf q}-\varepsilon_{\bf p}+l\Omega$. Absorption selects the negative of this energy and has the same squared local vertex. The factor $V$ in the definition of $W$ cancels the $1/V$ from the spatially integrated squared vertex. The scattering rates therefore read
\begin{align}
\begin{aligned}
&W_{{\bf q} \to {\bf p} } = {2\pi} |\chi_0|^2 \sum_{l=-\infty}^{\infty} \Phi^{(l)}_{{\bf q},{\bf p}} \, S(\varepsilon_{\bf q} - \varepsilon_{\bf p} + l\Omega).
\end{aligned}
\label{S-Wqp-0}
\end{align}
For box-normalized momentum states, the matrix element of one local projector scales as $V^{-1}$. Summing the independent local baths therefore gives the transition probability per unit time $W_{{\bf q}\to{\bf p}}/V$. In the thermodynamic limit, $V^{-1}\sum_{\bf q}\to\int[d{\bf q}]$, with $[d{\bf q}]=d^3q/(2\pi)^3$. Thus $W_{{\bf q}\to{\bf p}}[d{\bf p}]$ is a scattering probability per unit time into a momentum element, and $W$ has dimensions of volume per unit time. The condensate contribution is separated before this limit in SM Sec.~\ref{S-condensate-separation}; {the main-text stationary equation suppresses the common $1/V$ factor, which does not affect its solutions.} Summing the population equation over ${\bf p}$ gives zero after interchanging ${\bf p}$ and ${\bf q}$ in the loss term, verifying number conservation. At $\omega=0$, $S(\omega)$ is defined by its limiting product: for $\nu_B(\omega)\simeq c_1\omega$, one obtains $S(0)=c_1/\beta$, with $N_b(\omega)=(e^{\beta\omega}-1)^{-1}$.

\subsubsection{Pairwise detailed balance and its driven violation}
\label{S-pairwise-balance}
The thermal bath obeys the KMS condition $S(-\omega)=e^{-\beta\omega}S(\omega)$~\cite{kubo1957statistical,martin1959theory}, because $1+N_b(\omega)=e^{\beta\omega}N_b(\omega)$ for $\omega>0$. The Hermitian local-density vertex obeys $\Phi^{(-l)}_{{\bf p},{\bf q}}=\Phi^{(l)}_{{\bf q},{\bf p}}$. Hence an arbitrary connected pair has
\begin{align}
\begin{aligned}
&\frac{W^{(0)}_{{\bf q}\to{\bf p}}}{W^{(0)}_{{\bf p}\to{\bf q}}} =e^{\beta(\varepsilon_{\bf q}-\varepsilon_{\bf p})},\qquad \frac{f_{\bf p}^{\rm eq}}{1+f_{\bf p}^{\rm eq}} =e^{-\beta(\varepsilon_{\bf p}-\mu)}.
\end{aligned}
\label{S-pairwise-rate-ratio}
\end{align}
Their product verifies $f_{\bf q}^{\rm eq}W^{(0)}_{{\bf q}\to{\bf p}}(1+f_{\bf p}^{\rm eq})=f_{\bf p}^{\rm eq}W^{(0)}_{{\bf p}\to{\bf q}}(1+f_{\bf q}^{\rm eq})$ for every pair. The chemical potential cancels from this flux balance; it is zero in the condensed phase and fixed by density in the normal state. Without driving only $l=0$ remains, while retaining only $l=0$ at nonzero drive also preserves this relation because the overlap is reciprocal.

For the full scattering rate, set $\omega_l=\varepsilon_{\bf q}-\varepsilon_{\bf p}+l\Omega$ and $a_l=2\pi|\chi_0|^2\Phi^{(l)}_{{\bf q},{\bf p}}$. Pairing opposite harmonics in the reverse scattering rate gives
\begin{align}
\begin{aligned}
&W_{{\bf q}\to{\bf p}}=\sum_l a_lS(\omega_l),\qquad W_{{\bf p}\to{\bf q}}=\sum_l a_l e^{-\beta\omega_l}S(\omega_l).
\end{aligned}
\label{S-full-driven-pair}
\end{align}
The factors $e^{-\beta l\Omega}$ generally prevent the equilibrium ratio from being extracted from these sums. This is the distinction between pairwise detailed balance and the total steady-state gain--loss balance. The presence of the $l=0$ term inside the full sum does not restore detailed balance; exceptional cancellations of sidebands are not excluded.

\subsection{Separation of the Condensate in the Thermodynamic Limit}
\label{S-condensate-separation}

To account for the macroscopic occupation of the condensate mode at momentum ${\bf p}_0={\bf 0}$ in a Bose-Einstein condensate, we need to separate the condensate and excited-state occupations:
\begin{align}
\begin{aligned}
&f_{\bf p} = n_0 V \delta_{{\bf p},{\bf 0}} + f_{\mathrm{exc},{\bf p}},
\end{aligned}
\end{align}
where $n_0$ is the condensate number density (an intensive quantity), $V$ is the system volume (an extensive quantity), and $f_{\mathrm{exc},{\bf p}}$ is the occupation of an excited state, which remains finite at fixed nonzero momentum as $V\to\infty$. We set $f_{\mathrm{exc},{\bf0}}=0$ to avoid counting the condensate twice and write $f_{\bf p}$ for the excited-state occupation when ${\bf p}\ne0$.

To derive the Floquet-Boltzmann equation in the thermodynamic limit ($V \to \infty$), we start from the discrete form of the equation and explicitly separate the condensate contribution.

{\it Equation for the Condensate Mode ---}
For the condensate (${\bf p}={\bf 0}$), the transitions between ${\bf q}={\bf 0}$ to ${\bf p}={\bf 0}$ cancel. Only transitions involving excited states contribute:
\begin{align}
\begin{aligned}
&\sum_{{\bf q} \neq {\bf 0}} \left[ f_{\bf q} W_{{\bf q} \to {\bf 0}} (1+f_{{\bf 0}}) - f_{{\bf 0}} W_{{\bf 0} \to {\bf q}} (1+f_{\bf q}) \right] = 0.
\end{aligned}
\end{align}
We substitute $f_{\bf0}=n_0V$, divide this equation by $V^2$, and take the thermodynamic limit $V^{-1}\sum_{{\bf q}\ne{\bf0}}\to\int d^3q/(2\pi)^3$, using $(1+f_{\bf0})/V=n_0+V^{-1}\to n_0$. We obtain
\begin{align}
\begin{aligned}
&n_0 \int \frac{d^3 q}{(2\pi)^3} \left[ f_{\bf q} W_{{\bf q} \to {\bf 0}} - W_{{\bf 0} \to {\bf q}} (1+f_{\bf q}) \right] = 0.
\end{aligned}
\label{S-EftGS-0}
\end{align}
{\it Equation for Excited States --- }
For any excited state with momentum ${\bf p} \neq {\bf 0}$, the Floquet-Boltzmann equation describing the balance of particle currents is:
\begin{align}
\begin{aligned}
&\sum_{\bf q} \left[ f_{\bf q} W_{{\bf q} \to {\bf p}} (1+f_{\bf p}) - f_{\bf p} W_{{\bf p} \to {\bf q}} (1+f_{\bf q}) \right] = 0.
\end{aligned}
\end{align}
We split the sum over ${\bf q}$ into the condensate term (${\bf q}={\bf 0}$) and the sum over all other excited states (${\bf q} \neq {\bf 0}$):
\begin{align}
\begin{aligned}
&\left[ f_{{\bf 0}} W_{{\bf 0} \to {\bf p}} (1+f_{\bf p}) - f_{\bf p} W_{{\bf p} \to {\bf 0}} (1+f_{{\bf 0}}) \right] + \sum_{{\bf q} \neq {\bf 0}} \left[ f_{\bf q} W_{{\bf q} \to {\bf p}} (1+f_{\bf p}) - f_{\bf p} W_{{\bf p} \to {\bf q}} (1+f_{\bf q}) \right] = 0.
\end{aligned}
\end{align}
We again substitute $f_{\bf0}=n_0V$, divide this equation by $V$, and take the thermodynamic limit $V^{-1}\sum_{{\bf q}\ne{\bf0}}\to\int d^3q/(2\pi)^3$. Using $f_{\bf0}/V=n_0$ and $(1+f_{\bf0})/V=n_0+V^{-1}\to n_0$, we obtain
\begin{align}
\begin{aligned}
&n_0 \left[W_{{\bf 0} \to {\bf p}}(1 + f_{\bf p}) - f_{\bf p} W_{{\bf p} \to {\bf 0}}\right] + \int \frac{d^3 q}{(2\pi)^3} \left[ f_{\bf q} W_{{\bf q} \to {\bf p}}(1 + f_{\bf p}) - f_{\bf p} W_{{\bf p} \to {\bf q}}(1 + f_{\bf q}) \right] = 0,
\end{aligned}
\label{S-EfES-0}
\end{align}
Replacing the right-hand side of Eq.~\eqref{S-EfES-0} by $\partial_t f_{\bf p}$ gives the time-dependent, thermodynamic-limit version of the main-text Floquet-Boltzmann equation. The condensate term contains both gain and loss. Collecting the terms independent of and linear in $f_{\bf p}$ gives $0=I_{\bf p}[f]-\Gamma_{\bf p}[f]f_{\bf p}$, with
\begin{align}
\begin{aligned}
&I_{\bf p}[f]=n_0W_{0\to{\bf p}}+\int\frac{d^3q}{(2\pi)^3}f_{\bf q}W_{{\bf q}\to{\bf p}}, \\
&\Gamma_{\bf p}[f]=n_0(W_{{\bf p}\to0}-W_{0\to{\bf p}}) +\int\frac{d^3q}{(2\pi)^3}\left[W_{{\bf p}\to{\bf q}}(1+f_{\bf q})-f_{\bf q}W_{{\bf q}\to{\bf p}}\right].
\end{aligned}
\label{S-Ipf-Gammapf}
\end{align}
With $n_0$ and the other occupations held fixed, the kinetic equation is $\partial_t f_{\bf p}=I_{\bf p}-\Gamma_{\bf p}f_{\bf p}$, so we call $\Gamma_{\bf p}$ the net relaxation rate. For $\Gamma_{\bf p}>0$, the occupation relaxes toward $I_{\bf p}/\Gamma_{\bf p}$. The bath spectral step can enter both numerator and denominator.

The condensate contribution in Eq.~\eqref{S-EfES-0} contains no integration
over the initial system momentum. The condensate channel therefore inherits the bath spectral step directly,
whereas the corresponding excited-state contribution is integrated over the available phase space.

{\it Particle Density Conservation --- }
The above two integral equations are supplemented by a third constraint for the conservation of the total particle density, $n_\mathrm{tot}$. Starting from the total particle number $N_\mathrm{tot} = f_{\bf 0} + \sum_{{\bf p}\neq{\bf 0}} f_{\bf p}$, we divide by the volume $V$ and take the thermodynamic limit:
\begin{align}
\begin{aligned}
&n_\mathrm{tot} = \frac{N_\mathrm{tot}}{V} = \frac{f_{\bf 0}}{V} + \frac{1}{V}\sum_{{\bf p}\neq{\bf 0}} f_{\bf p} \quad \xrightarrow{V\to\infty} \quad n_\mathrm{tot} = n_0 + \int \frac{d^3 p}{(2\pi)^3} \, f_{\bf p}.
\end{aligned}
\label{S-PDC-0}
\end{align}
These equations [Eqs.~\eqref{S-EfES-0}, \eqref{S-EftGS-0}, and \eqref{S-PDC-0}] specify the steady-state problem, including the condensate density $n_0$. Their consistency follows directly by integrating Eq.~\eqref{S-EfES-0} over ${\bf p}$: the terms describing scattering between excited states cancel under ${\bf p}\leftrightarrow{\bf q}$, and the remaining condensate term is the negative of Eq.~\eqref{S-EftGS-0}. Thus the condensate equation is a particle-balance condition, while the fixed total density selects the physical solution.

\subsection{Scattering Rates without Magnetic Field}
\label{S-parabolic-rates}

For a local, isotropic scattering potential where the bath couples to the particle density, the total non-equilibrium scattering rates established in the previous section read [Eq.~\eqref{S-Wqp-0}]:
\begin{align}
\begin{aligned}
&W_{{\bf q} \to {\bf p}} = {2\pi} \sum_{l=-\infty}^{\infty} |\chi_{{\bf q}, {\bf p}}(l)|^2 \, S(\varepsilon_{\bf q} - \varepsilon_{{\bf p}} + l\Omega),
\end{aligned}
\end{align}
where the coupling-weighted Floquet overlap is $|\chi_{{\bf q}, {\bf p}}(l)|^2=|\chi_0|^2\Phi^{(l)}_{{\bf q},{\bf p}}$. It is the squared modulus of the $l$-th Fourier harmonic of $\chi_{{\bf q},{\bf p}}(t)$, where
\begin{align}
\begin{aligned}
&\chi_{{\bf q}, {\bf p}}(t) = \chi_0~\phi_{\bf p}^*(t) \phi_{\bf q}^{}(t) = \sum_{l=-\infty}^{\infty} \chi_{{\bf q}, {\bf p}}(l) e^{-il\Omega t}.
\end{aligned}
\end{align}
Thus $\chi_{{\bf q},{\bf p}}(l)=T^{-1}\int_0^Tdt\,e^{il\Omega t}\chi_{{\bf q},{\bf p}}(t)$, with $\chi_0$ the local coupling amplitude.

We consider a uniform three-dimensional Bose gas with an isotropic parabolic dispersion. In the absence of a static magnetic field, the system is driven by a spatially uniform, time-periodic vector potential with zero time average, ${\bf A}(t)$. The single-particle Hamiltonian is:
\begin{align}
\begin{aligned}
&\hat h(t) = \frac{1}{2m} \big(\hat{\bf p} - {\bf A}(t)\big)^2 = \frac{\hat{\bf p}^2}{2m} - \frac{1}{m} \hat{\bf p} \cdot {\bf A}(t) + \frac{1}{2m} {\bf A}^2(t).
\end{aligned}
\end{align}
Because the spatially uniform drive preserves continuous translational symmetry, the solutions to the time-dependent Schr\"odinger equation are plane waves modulated by a periodic Floquet envelope:
\begin{align}
\begin{aligned}
&\langle {\bf r} | \psi_{\bf p}(t) \rangle = \frac{1}{\sqrt{V}} e^{i {\bf p} \cdot {\bf r}} e^{-i \varepsilon_{\bf p} t} \phi_{\bf p}(t).
\end{aligned}
\end{align}
The Floquet quasienergy is given by the cycle-averaged kinetic energy, $\varepsilon_{\bf p} = {\bf p}^2/2m + \langle {\bf A}^2 \rangle_T / 2m$. The $T$-periodic envelope $\phi_{\bf p}(t)$ integrates the purely oscillatory components of the Hamiltonian:
\begin{align}
\begin{aligned}
&\phi_{\bf p}(t) = \exp\left[-i \int_0^t d\tau \left( -\frac{1}{m} {\bf p} \cdot {\bf A}(\tau) + \frac{1}{2m} \Big({\bf A}^2(\tau) - \langle {\bf A}^2 \rangle_T\Big) \right) \right].
\end{aligned}
\end{align}
When evaluating the time-dependent envelope overlap $\chi_{{\bf q}, {\bf p}}(t)$ for a transition from an initial momentum ${\bf q}$ to a final momentum ${\bf p}$, the diamagnetic phase terms proportional to ${\bf A}^2(\tau)$ and $\langle {\bf A}^2 \rangle_T$ cancel between the bra and the ket because they are spatially uniform and independent of momentum. The time-dependent Floquet overlap is then:
\begin{align}
\begin{aligned}
&\chi_{{\bf q}, {\bf p}}(t) = \chi_0~ \exp\left[ i \frac{1}{m} ({\bf q} - {\bf p}) \cdot \int_0^t d\tau \, {\bf A}(\tau) \right] = \chi_0~ \exp\left[ i \frac{1}{m} {\bf k}_\perp \cdot \int_0^t d\tau \, {\bf A}(\tau) \right],
\end{aligned}
\end{align}
where ${\bf k}_\perp \equiv {\bf q}_\perp - {\bf p}_\perp$ is the in-plane difference between the initial and final system momenta. This choice fixes the sign of the overlap phase. Note that longitudinal momentum transfer $q_z - p_z$ inherently decouples from the in-plane drive.

We now specialize to a circularly polarized drive in the $xy$-plane, parameterized by ${\bf A}(t) = A_0 (\cos \Omega t, \sin \Omega t, 0)$. The exact time integral of the vector potential yields:
\begin{align}
\begin{aligned}
&\int_0^t d\tau \, {\bf A}(\tau) = \frac{A_0}{\Omega} (\sin \Omega t, 1 - \cos \Omega t, 0).
\end{aligned}
\end{align}
Expressing the in-plane momentum transfer vector in polar coordinates as ${\bf k}_\perp = (k_\perp \cos \theta_\perp, k_\perp \sin \theta_\perp, 0)$, the dot product evaluates directly to:
\begin{align}
\begin{aligned}
&i \frac{1}{m} {\bf k}_\perp \cdot \int_0^t d\tau \, {\bf A}(\tau) = i \frac{A_0 k_\perp}{m \Omega} \big( \cos \theta_\perp \sin \Omega t + \sin \theta_\perp (1 - \cos \Omega t) \big) = i \lambda_0 k_\perp \big[ \sin(\Omega t - \theta_\perp) + \sin \theta_\perp \big],
\end{aligned}
\end{align}
where we have introduced the characteristic length $\lambda_0 \equiv A_0/(m\Omega)$; the product $\lambda_0 k_\perp$ is dimensionless. 

Substituting this into the overlap function isolates the dynamical periodic phase from a static geometric phase shift:
\begin{align}
\begin{aligned}
&\chi_{{\bf q}, {\bf p}}(t) = \chi_0 ~e^{i \lambda_0 k_\perp \sin \theta_\perp} \exp\left[ i \lambda_0 k_\perp \sin(\Omega t - \theta_\perp) \right].
\end{aligned}
\end{align}
The global, time-independent prefactor $e^{i \lambda_0 k_\perp \sin \theta_\perp} \equiv e^{i\mathcal{C}}$ is an irrelevant gauge phase that drops out completely upon taking the squared modulus. To extract the exact $l$-th Fourier harmonic required by the golden rule, we expand the periodic dynamical phase using the Jacobi-Anger identity,
\begin{align}
\begin{aligned}
&e^{i z \sin \alpha} = \sum_{k=-\infty}^\infty J_k(z) e^{i k \alpha} \to \chi_{{\bf q}, {\bf p}}(t) = \chi_0 e^{i\mathcal{C}} \sum_{k=-\infty}^\infty J_k(\lambda_0 k_\perp) e^{i k (\Omega t - \theta_\perp)}.
\end{aligned}
\end{align}
By direct formal comparison with the defining discrete Fourier series $\chi_{{\bf q}, {\bf p}}(t) = \sum_{l=-\infty}^\infty \chi_{{\bf q}, {\bf p}}(l) e^{-i l \Omega t}$, we identify the $l$-th Floquet harmonic by isolating the index $k = -l$:
\begin{align}
\begin{aligned}
&\chi_{{\bf q}, {\bf p}}(l) = \chi_0 ~ e^{i\mathcal{C}} J_{-l}(\lambda_0 k_\perp) e^{i l \theta_\perp}, \qquad \big| \chi_{{\bf q}, {\bf p}}(l) \big|^2 = |\chi_0|^2 ~ J_l^2(\lambda_0 k_\perp).
\end{aligned}
\end{align}
Here we used $J_{-l}(x)=(-1)^lJ_l(x)$. The prefactor is linear in $\chi_0$ at the amplitude level and quadratic only after taking its squared modulus.

This derivation shows that under circular polarization, the strength of the drive-assisted scattering processes depends only on the scalar magnitude of the in-plane momentum transferred to the bath $k_\perp = |{\bf p}_\perp - {\bf q}_\perp| = \sqrt{p_\perp^2 + q_\perp^2 - 2 p_\perp q_\perp \cos(\phi_p - \phi_q)}$. Substituting this derived overlap back into Eq.~\eqref{S-Wqp-0}, the scattering rates are:
\begin{align}
\begin{aligned}
&W_{{\bf q} \to {\bf p} } = {2\pi} |\chi_0|^2 \sum_{l=-\infty}^{\infty} J_l^2\left(\frac{A_0}{m\Omega} |{\bf p}_\perp - {\bf q}_\perp|\right) \, S(\varepsilon_{\bf q} - \varepsilon_{\bf p} + l\Omega).
\end{aligned}
\end{align}
The common shift $\langle{\bf A}^2\rangle_T/(2m)$ cancels from $\varepsilon_{\bf q}-\varepsilon_{\bf p}$. In the following angular and energy projections, we therefore measure quasienergies relative to the band bottom and write $\varepsilon_{\bf p}={\bf p}^2/(2m)$.

The scattering rates derived above depend on the azimuthal angles of the initial and final momenta only through their relative angle $\phi \equiv \phi_p - \phi_q$. Therefore we expand both the occupation function and the scattering rates into azimuthal Fourier series. Since the state depends on $p_\perp, p_z$, and $\phi_p$, we denote $p \equiv (p_\perp, p_z)$ to compress the indices:
\begin{align}
\begin{aligned}
&f({\bf p}) = \sum_{m=-\infty}^{\infty} f_m(p) e^{i m \phi_p},\qquad \bar{f}({\bf p}) \equiv 1 + f({\bf p}) = \sum_{m=-\infty}^{\infty} \bar{f}_m(p) e^{i m \phi_p},\qquad \bar{f}_m(p) = \delta_{m,0} + f_m(p), \\
&W_{{\bf q} \to {\bf p}} = \sum_{n=-\infty}^{\infty} W_n(q, p) e^{in(\phi_p - \phi_q)}.
\end{aligned}
\end{align}
Because the scattering rates are purely real and symmetric with respect to angle reversal ($\phi \to -\phi$), their Fourier coefficients are real and symmetric in the index: $W_{-n}(q, p) = W_n^*(q, p) = W_n(q, p)$.

Substituting these expansions into the Floquet-Boltzmann collision integral, we extract the equation governing the $m$-th azimuthal harmonic. For the scattering-in flux, integrating out the initial azimuthal angle $\phi_q$ yields:
\begin{align}
\begin{aligned}
&\int_0^{2\pi} \frac{d\phi_q}{2\pi} f_{\bf q} W_{{\bf q} \to {\bf p}} \bar{f}_{\bf p} = \int_0^{2\pi} \frac{d\phi_q}{2\pi} \left( \sum_{m'} f_{m'}(q) e^{i m' \phi_q} \right) \left( \sum_n W_n(q, p) e^{i n (\phi_p - \phi_q)} \right) \left( \sum_k \bar{f}_k(p) e^{i k \phi_p} \right) \\
&\quad = \sum_{n, k} f_n(q) W_n(q, p) \bar{f}_k(p) e^{i (n+k) \phi_p} = \sum_m e^{i m \phi_p} \sum_n f_n(q) W_n(q, p) \bar{f}_{m-n}(p),
\end{aligned}
\end{align}
where we grouped the terms into the $m$-th total angular harmonic by shifting the dummy index $k = m-n$. Applying this procedure to the scattering-out flux, and utilizing the symmetry $W_{-n}(p, q) = W_n(p, q)$, we obtain:
\begin{align}
\begin{aligned}
&\int_0^{2\pi} \frac{d\phi_q}{2\pi} f_{\bf p} W_{{\bf p} \to {\bf q}} \bar{f}_{\bf q} = \int_0^{2\pi} \frac{d\phi_q}{2\pi} \left( \sum_{k} f_{k}(p) e^{i k \phi_p} \right) \left( \sum_n W_n(p, q) e^{i n (\phi_q - \phi_p)} \right) \left( \sum_{m'} \bar{f}_{m'}(q) e^{i m' \phi_q} \right) \\
&\quad = \sum_{n, k} f_k(p) W_n(p, q) \bar{f}_{-n}(q) e^{i (k-n) \phi_p} = \sum_m e^{i m \phi_p} \sum_n f_{m-n}(p) W_n(p, q) \bar{f}_{n}(q).
\end{aligned}
\end{align}
The second equality follows from $m'=-n$; in the final equality we relabel $n\to-n$ and use $W_{-n}=W_n$. The condensate contribution to the $m$-th harmonic is
\begin{align}
\begin{aligned}
&I_m^{\mathrm{cond}}(p) =n_0\left[ \delta_{m0}W_{{\bf0}\to{\bf p}} +f_m(p)(W_{{\bf0}\to{\bf p}}-W_{{\bf p}\to{\bf0}}) \right],
\end{aligned}
\end{align}
because the condensate-scattering rates have no azimuthal dependence.
The full collision integral thus decomposes into an orthogonal Fourier series $I({\bf p}) = \sum_m I_m(p) e^{im\phi_p} = 0$. By taking the difference between the in and out terms, the equation governing the $m$-th harmonic reads:
\begin{align}
\begin{aligned}
&I_m(p) = I_m^{\mathrm{cond}}(p) + \sum_{q} \sum_{n=-\infty}^{\infty} \Big[ f_n(q) W_n(q, p) \bar{f}_{m-n}(p) - f_{m-n}(p) W_n(p, q) \bar{f}_n(q) \Big] = 0,
\end{aligned}
\end{align}
where $\sum_q \equiv \int \frac{q_\perp d q_\perp d q_z}{(2\pi)^2}$ denotes integration over the remaining degrees of freedom. To expose whether the angular harmonics couple, we substitute $\bar{f}_k = \delta_{k,0} + f_k$ and distribute the Kronecker deltas:
\begin{align}
\begin{aligned}
&I_m(p) = I_m^{\mathrm{cond}}(p) \\
&\quad + \sum_{q} \left\{f_m(q) W_m(q, p) - f_m(p) W_0(p, q) + \sum_{n=-\infty}^{\infty} f_n(q) f_{m-n}(p) \Big[ W_n(q, p) - W_n(p, q) \Big] \right\} = 0.
\end{aligned}
\end{align}
This reveals that azimuthal harmonics do \textit{not} decouple. The cross-coupling coefficient vanishes when the scattering rates satisfy $W_n(q, p) = W_n(p, q)$, which is not valid in general.

However, because the circularly polarized drive and the uniform bath possess no preferred in-plane direction, we can test if an azimuthally isotropic steady-state ansatz $f({\bf p}) = f_0(p)$ (i.e., $f_m = 0$ for all $m \neq 0$) is a valid solution. Substituting this ansatz directly into the harmonic equations shows that the azimuthally isotropic sector is closed under the collision equation:
\begin{itemize}
    \item For $m \neq 0$: The linear terms $f_m(q)$ and $f_m(p)$ vanish. In the non-linear summation $\sum_n f_n(q) f_{m-n}(p)$, the product is non-zero if and only if both $n = 0$ and $m - n = 0$, which requires $m=0$. Thus, the summation and $I_m^{\mathrm{cond}}$ are zero for any $m \neq 0$, and the equation $I_{m \neq 0}(p) = 0$ is satisfied.
    \item For $m = 0$: The condition isolates the $n=0$ term in the summation. Combining it with the linear and condensate terms restores the azimuthally averaged Floquet-Boltzmann equation:
\end{itemize}
\begin{align}
\begin{aligned}
&I_0(p)= I_0^{\mathrm{cond}}(p) + \sum_{q} \Big[ f_0(q) W_0(q, p) \big(1 + f_0(p)\big) - f_0(p) W_0(p, q) \big(1 + f_0(q)\big) \Big] = 0.
\end{aligned}
\end{align}
The above analysis establishes that cylindrical symmetry, $f({\bf p})=f(p_\perp,p_z)$, is a consistent ansatz and allows us to drop the azimuthal index within this sector. It does not establish uniqueness or stability against perturbations outside this sector. Scattering between excited states is described by the azimuthally averaged scattering rates $\overline{W}_{{\bf q} \to {\bf p}} \equiv W_0(q, p) = \int_0^{2\pi} \frac{d\phi}{2\pi} W_{{\bf q} \to {\bf p}}(\phi)$:
\begin{align}
\begin{aligned}
&\overline{W}_{{\bf q} \to {\bf p}} = {2\pi} |\chi_0|^2 \sum_{l=-\infty}^{\infty} \left[ \int_0^{2\pi} \frac{d\phi}{2\pi} J_l^2\left(\frac{A_0}{m\Omega} \sqrt{p_\perp^2 + q_\perp^2 - 2 p_\perp q_\perp \cos\phi}\right) \right] S(\varepsilon_{\bf q} - \varepsilon_{\bf p} + l\Omega).
\end{aligned}
\end{align}
We can apply Graf's addition theorem for Bessel functions to evaluate this angular integration. Define $z e^{i\psi}=x-y e^{-i\phi}$, so $z=\sqrt{x^2+y^2-2xy\cos\phi}$. The identity follows by expanding both factors in
\begin{align}
\begin{aligned}
&e^{-ix\sin t}e^{iy\sin(t+\phi)} =\sum_{a,b}J_a(x)J_b(y)e^{-i(a-b)t}e^{ib\phi} =e^{-iz\sin(t-\psi)}
\end{aligned}
\end{align}
and equating the coefficients of $e^{-ilt}$:
\begin{align}
\begin{aligned}
&J_l\left(\sqrt{x^2+y^2-2 x y \cos \phi}\right) e^{i l \psi}=\sum_{k=-\infty}^{\infty} J_{l+k}(x) J_k(y) e^{i k \phi}.
\end{aligned}
\end{align}
Taking the squared modulus and using
$\int_0^{2\pi}d\phi\,e^{i(k-k')\phi}/(2\pi)=\delta_{kk'}$
gives $\int_0^{2\pi}d\phi\,J_l^2(z)/(2\pi)=\sum_kJ_{l+k}^2(x)J_k^2(y)$.
Relabeling $k=-l-l'$ and using $J_{-n}^2=J_n^2$ gives the form below.
Here $p=|{\bf p}|$ and $q=|{\bf q}|$. We define $x \equiv \lambda_p \sin\theta_p$ and $y \equiv \lambda_q \sin\theta_q$, where $\lambda_{p} \equiv p A_0/m\Omega$, integrate out the relative azimuthal angle $\phi$, and obtain
\begin{align}
\begin{aligned}
&\int_0^{2\pi} \frac{d\phi}{2\pi} J_l^2\left(\frac{A_0}{m\Omega} \sqrt{p_\perp^2 + q_\perp^2 - 2 p_\perp q_\perp \cos\phi}\right) = \sum_{l'=-\infty}^{\infty} J_{l'}^2(\lambda_p \sin\theta_p) J_{l'+l}^2(\lambda_q \sin\theta_q).
\end{aligned}
\label{angle-averaged-rates}
\end{align}
\subsection{Projected Floquet-Boltzmann Equations}
\label{Projected-FBE}

The circularly polarized drive in the $xy$-plane breaks the 3D rotational symmetry of the system, reducing it to a cylindrical symmetry about the $z$-axis. Within the cylindrically symmetric sector considered here, the non-equilibrium steady-state occupation function $f_{\bf p}$ depends on the momentum magnitude $p = |{\bf p}|$ and the polar angle $\theta_p$ relative to the $z$-axis, but is independent of the azimuthal angle $\phi$. Furthermore, the symmetry of the parabolic dispersion and the in-plane drive under reflection $z \to -z$ allows us to choose the reflection-even sector, $f(p,\theta_p)=f(p,\pi-\theta_p)$. As above, this is a symmetry-consistent ansatz.

To solve the Floquet-Boltzmann equation efficiently, we decompose the angular dependence of the occupation function into Legendre polynomials $P_n(\cos\theta)$. For this reflection-even sector, only nonnegative even orders contribute:
\begin{align}
\begin{aligned}
&f(p, \theta) = \sum_{n \in \text{even}} x_n(p) P_n(\cos\theta_p) = \sum_{n \in \text{even}} x_n(\varepsilon_p) P_n(\cos\theta_p)
\end{aligned}
\label{eq:legendre_ansatz}
\end{align}
The coefficient normalization follows from Legendre orthogonality:
\begin{align}
\begin{aligned}
&\int_{-1}^{1}du\,P_n(u)P_k(u)=\frac{2\delta_{nk}}{2n+1}, \qquad x_n(\varepsilon)=\frac{2n+1}{2}\int_{-1}^{1}du\,P_n(u)f(\varepsilon,u), \quad u=\cos\theta.
\end{aligned}
\label{S-Legendre-orthogonality}
\end{align}
In particular, $x_0$ is the angle-averaged occupation. All Legendre sums below run over nonnegative even orders.
This expansion transforms the high-dimensional integral equation into a coupled system of 1D integral equations for the radial coefficients $x_n(\varepsilon_p)$.
We define the projection onto the $k$-th Legendre mode at energy $\varepsilon$ as:
\begin{align}
\begin{aligned}
&\langle {\cal O} \rangle_{k, \varepsilon} \equiv \int \frac{d^3{\bf p}}{(2\pi)^3} \delta(\varepsilon - \varepsilon_p) P_k(\cos\theta_p) {\cal O}({\bf p})
\end{aligned}
\end{align}
This is an unnormalized projection. A factor $(2k+1)/2$ is needed to extract a Legendre coefficient, but multiplying an equation equal to zero by that factor does not change its solutions.

{\it Equation for Condensate} $(p=0)$:
Starting from the equation for the condensate mode~[Eq.~\eqref{S-EftGS-0}]:
\begin{align}
\begin{aligned}
&n_0 \int \frac{d^3{\bf q}}{(2\pi)^3} \left[ f_{{\bf q}} W_{{\bf q} \to {\bf 0}} - W_{{\bf 0} \to {\bf q}}(1+f_{{\bf q}}) \right] = 0
\end{aligned}
\end{align}
We substitute the ansatz for $f_{{\bf q}}$ and insert the spectral identity $1 = \int d\varepsilon \delta(\varepsilon - \varepsilon_q)$ to group terms by energy:
\begin{align}
\begin{aligned}
&n_0 \int d\varepsilon \sum_{n} \left[ x_n(\varepsilon) {\cal A}_{\text{fwd}}^{(n)}(\varepsilon) - (\delta_{n,0} + x_n(\varepsilon)) {\cal A}_{\text{bwd}}^{(n)}(\varepsilon) \right] = 0
\end{aligned}
\end{align}
For $n_0>0$, the integral must vanish; for $n_0=0$, the equation is identically satisfied. This defines the energy-dependent Legendre projections ${\cal A}$ of the scattering rates:
\begin{align}
\begin{aligned}
&{\cal A}_{\text{fwd}}^{(n)}(\varepsilon) = \int \frac{d^3{\bf q}}{(2\pi)^3} \delta(\varepsilon - \varepsilon_q) P_n(\cos\theta_q) W_{{\bf q} \to {\bf 0}}\qquad {\cal A}_{\text{bwd}}^{(n)}(\varepsilon) = \int \frac{d^3{\bf q}}{(2\pi)^3} \delta(\varepsilon - \varepsilon_q) P_n(\cos\theta_q) W_{{\bf 0} \to {\bf q}}
\end{aligned}
\end{align}
{\it Equation for Excited States} $(p \neq 0)$:
The Floquet-Boltzmann equation for excited states is [Eq.~\eqref{S-EfES-0}]:
\begin{align}
\begin{aligned}
&n_0 \left[ W_{{\bf 0} \to {\bf p}}(1+f_{{\bf p}}) - f_{{\bf p}} W_{{\bf p} \to {\bf 0}} \right] + \int \frac{d^3{\bf q}}{(2\pi)^3} \left[ f_{{\bf q}} W_{{\bf q} \to {\bf p}} (1+f_{{\bf p}}) - f_{{\bf p}} W_{{\bf p} \to {\bf q}} (1+f_{{\bf q}}) \right] = 0.
\end{aligned}
\end{align}
We substitute the ansatz for $f_{{\bf p}}$ and $f_{{\bf q}}$, and project the Floquet-Boltzmann equation for excited states onto the $k$-th Legendre mode at energy $\varepsilon$ by applying $\langle \dots \rangle_{k, \varepsilon} = \int \frac{d^3{\bf p}}{(2\pi)^3} \delta(\varepsilon - \varepsilon_p) P_k(\cos\theta_p)$ to the entire equation. The projected equation for the coefficients $x_k(\varepsilon)$ is:
\begin{align}
\begin{aligned}
&n_0 \sum_{n} \left[ (\delta_{n,0} + x_n(\varepsilon)) {\cal B}_{\text{bwd}}^{(n, k)}(\varepsilon) - x_n(\varepsilon) {\cal B}_{\text{fwd}}^{(n, k)}(\varepsilon) \right] \\
&\quad + \sum_{n, m} \int d\varepsilon' \Bigg[ (\delta_{n,0} + x_n(\varepsilon)) x_m(\varepsilon') {\cal C}_{\text{fwd}}^{(n,m,k)}(\varepsilon', \varepsilon) - x_n(\varepsilon) (\delta_{m,0} + x_m(\varepsilon')) {\cal C}_{\text{bwd}}^{(n,m,k)}(\varepsilon, \varepsilon') \Bigg] = 0,
\end{aligned}
\end{align}
in which we defined the following Legendre projections of the scattering rates:
\begin{align}
\begin{aligned}
&{\cal B}_{\text{bwd}}^{(n, k)}(\varepsilon) = \int \frac{d^3{\bf p}}{(2\pi)^3} \delta(\varepsilon - \varepsilon_p) P_k(\cos\theta_p) P_n(\cos\theta_p) W_{{\bf 0} \to {\bf p}} \\
&{\cal B}_{\text{fwd}}^{(n, k)}(\varepsilon) = \int \frac{d^3{\bf p}}{(2\pi)^3} \delta(\varepsilon - \varepsilon_p) P_k(\cos\theta_p) P_n(\cos\theta_p) W_{{\bf p} \to {\bf 0}}
\end{aligned}
\end{align}
\begin{align}
\begin{aligned}
&{\cal C}_{\text{fwd}}^{(n,m,k)}(\varepsilon', \varepsilon) = \int \frac{d^3{\bf p}}{(2\pi)^3} \int \frac{d^3{\bf q}}{(2\pi)^3} \delta(\varepsilon - \varepsilon_p) \delta(\varepsilon' - \varepsilon_q) \times P_k(\cos\theta_p) P_n(\cos\theta_p) P_m(\cos\theta_q) W_{{\bf q} \to {\bf p}} \\
&{\cal C}_{\text{bwd}}^{(n,m,k)}(\varepsilon, \varepsilon') = \int \frac{d^3{\bf p}}{(2\pi)^3} \int \frac{d^3{\bf q}}{(2\pi)^3} \delta(\varepsilon - \varepsilon_p) \delta(\varepsilon' - \varepsilon_q) \times P_k(\cos\theta_p) P_n(\cos\theta_p) P_m(\cos\theta_q) W_{{\bf p} \to {\bf q}}
\end{aligned}
\end{align}
{\it Equation for Particle Conservation}:
Starting from the particle conservation in momentum space [Eq.~\eqref{S-PDC-0}]:
\begin{align}
\begin{aligned}
&n_{\mathrm{tot}} = n_0 + \int \frac{d^3{\bf p}}{(2\pi)^3} f_{{\bf p}},
\end{aligned}
\end{align}
we substitute the Legendre expansion for the occupation function, $f_{{\bf p}} = \sum_n x_n(\varepsilon_p) P_n(\cos\theta_p)$. We transform the integration measure to the energy representation by inserting the spectral identity $1 = \int_0^\infty d\varepsilon \, \delta(\varepsilon - \varepsilon_p)$:
\begin{align}
\begin{aligned}
&n_{\mathrm{tot}} = n_0 + \int_0^\infty d\varepsilon \sum_n x_n(\varepsilon) \left[ \int \frac{d^3{\bf p}}{(2\pi)^3} \delta(\varepsilon - \varepsilon_p) P_n(\cos\theta_p) \right].
\end{aligned}
\end{align}
Using the density-of-states weight ${\cal D}(\varepsilon)$ defined later in Eq.~\eqref{eq:DOS_def}, the momentum integral reduces to an angular average over the energy shell:
\begin{align}
\begin{aligned}
&\int \frac{d^3{\bf p}}{(2\pi)^3} \delta(\varepsilon - \varepsilon_p) P_n(\cos\theta_p) = {\cal D}(\varepsilon) \int_{-1}^{1} d(\cos\theta) P_n(\cos\theta).
\end{aligned}
\end{align}
Due to the orthogonality of the Legendre polynomials, the angular integral is $\int_{-1}^{1} d(\cos\theta) P_n(\cos\theta) = 2 \delta_{n,0}$. Consequently, only the isotropic ($n=0$) mode of the excited-state occupation contributes to the total density. The explicit conservation equation is:
\begin{align}
\begin{aligned}
&n_{\mathrm{tot}} = n_0 + 2 \int_0^\infty d\varepsilon \, {\cal D}(\varepsilon) x_0(\varepsilon).
\end{aligned}
\label{eq:conservation_energy}
\end{align}
This equation shows that the total particle number is partitioned between the condensate and the energy-integrated, angle-averaged population of the excited states.

{\it Explicit Evaluation of the Scattering Rate Projections}:
Here we explicitly integrate out the Dirac delta functions appearing in the definitions of these projections.
We utilize the isotropic parabolic dispersion $\varepsilon_p = p^2/2m$, which defines the momentum magnitude at energy $\varepsilon$ as $p_\varepsilon = \sqrt{2m\varepsilon}$. For $\varepsilon>0$ and $p\geq0$, the radial delta function is
\begin{align}
\begin{aligned}
&\delta\!\left(\varepsilon-\frac{p^2}{2m}\right) =\frac{m}{p_\varepsilon}\delta(p-p_\varepsilon), \qquad \int_0^\infty dp\,p^2\delta\!\left(\varepsilon-\frac{p^2}{2m}\right) =mp_\varepsilon.
\end{aligned}
\end{align}
The projection of any cylindrically symmetric function $F({\bf p}) \equiv F(p, \theta)$ onto an energy shell transforms the momentum integration measure as follows:
\begin{align}
\begin{aligned}
&\int \frac{d^3{\bf p}}{(2\pi)^3} \delta(\varepsilon - \varepsilon_p) F({\bf p}) = {\cal D}(\varepsilon) \int_{-1}^{1} d(\cos\theta) \, F(p_\varepsilon, \theta),
\end{aligned}
\end{align}
where the density-of-states weight per unit $d(\cos\theta)$, after the azimuthal integral, is
\begin{align}
\begin{aligned}
&{\cal D}(\varepsilon) \equiv \frac{m \sqrt{2m\varepsilon}}{(2\pi)^2}.
\end{aligned}
\label{eq:DOS_def}
\end{align}
The full three-dimensional density of states is $2{\cal D}(\varepsilon)$; this accounts for the factor of two in Eq.~\eqref{eq:conservation_energy}.
Substituting the explicit scattering rates $W_{{\bf q}\to {\bf p}}$ into the definitions of these projections and defining the energy-dependent dimensionless drive amplitude
\begin{align}
\begin{aligned}
&\lambda_\varepsilon \equiv \frac{A_0 p_\varepsilon}{m\Omega},
\end{aligned}
\end{align} we obtain the following expressions: 
\begin{align}
\begin{aligned}
&{\cal A}_{\text{fwd}}^{(n)}(\varepsilon) = {2\pi}|\chi_0|^2 {\cal D}(\varepsilon) \sum_{l} S(\varepsilon + l\Omega) \int_{-1}^{1} d(\cos\theta) P_n(\cos\theta) J_l^2(\lambda_\varepsilon \sin\theta), \\
&{\cal A}_{\text{bwd}}^{(n)}(\varepsilon) = {2\pi}|\chi_0|^2 {\cal D}(\varepsilon) \sum_{l} S(-\varepsilon + l\Omega) \int_{-1}^{1} d(\cos\theta) P_n(\cos\theta) J_l^2(\lambda_\varepsilon \sin\theta).
\end{aligned}
\end{align}
\begin{align}
\begin{aligned}
&{\cal B}_{\text{bwd}}^{(n, k)}(\varepsilon) = {2\pi}|\chi_0|^2 {\cal D}(\varepsilon) \sum_{l} S(l\Omega - \varepsilon) \int_{-1}^{1} d(\cos\theta) P_k(\cos\theta) P_n(\cos\theta) J_l^2(\lambda_\varepsilon \sin\theta), \\
&{\cal B}_{\text{fwd}}^{(n, k)}(\varepsilon) = {2\pi}|\chi_0|^2 {\cal D}(\varepsilon) \sum_{l} S(l\Omega + \varepsilon) \int_{-1}^{1} d(\cos\theta) P_k(\cos\theta) P_n(\cos\theta) J_l^2(\lambda_\varepsilon \sin\theta).
\end{aligned}
\end{align}
For the Legendre projections ${\cal C}_{\mathrm{fwd}}$ and ${\cal C}_{\mathrm{bwd}}$ describing scattering between excited states, the Legendre factors do not depend on azimuth. The azimuthal integrations therefore replace $W$ by its angular average. Eq.~\eqref{angle-averaged-rates} then separates the remaining polar integrals:
\begin{align}
\begin{aligned}
&{\cal C}_{\text{fwd}}^{(n,m,k)}(\varepsilon', \varepsilon) = {2\pi}|\chi_0|^2 {\cal D}(\varepsilon) {\cal D}(\varepsilon') \sum_{l, l'} S(\varepsilon' - \varepsilon + l\Omega) \\
&\quad \times \left[ \int_{-1}^{1} d(\cos\theta_p) P_k(\cos\theta_p) P_n(\cos\theta_p) J_{l'}^2(\lambda_\varepsilon \sin\theta_p) \right] \times \left[ \int_{-1}^{1} d(\cos\theta_q) P_m(\cos\theta_q) J_{l+l'}^2(\lambda_{\varepsilon'} \sin\theta_q) \right].
\end{aligned}
\end{align}
\begin{align}
\begin{aligned}
&{\cal C}_{\text{bwd}}^{(n,m,k)}(\varepsilon, \varepsilon') = {2\pi}|\chi_0|^2 {\cal D}(\varepsilon) {\cal D}(\varepsilon') \sum_{l, l'} S(\varepsilon - \varepsilon' + l\Omega) \\
&\quad \times \left[ \int_{-1}^{1} d(\cos\theta_p) P_k(\cos\theta_p) P_n(\cos\theta_p) J_{l+l'}^2(\lambda_\varepsilon \sin\theta_p) \right] \times \left[ \int_{-1}^{1} d(\cos\theta_q) P_m(\cos\theta_q) J_{l'}^2(\lambda_{\varepsilon'} \sin\theta_q) \right].
\end{aligned}
\end{align}
The condensate map in Fig.~\ref{fig1}(a) samples a $50\times50$ grid in bath temperature and drive amplitude, with $n_0$ and the excited-state occupations determined self-consistently at each point.

\subsection{bosonic Floquet Fermi surfaces for the gapless bath}
\label{S-ohmic-nonanalyticities}
We first set $\nu_g=0$ while retaining the gapless spectrum $\nu_B(\omega>0)=c_1\omega+c_2\omega^2$, with $c_1>0$ and $c_2>0$. At $T_B>0$, the bath function is
\begin{align}
\begin{aligned}
&S(\omega)=\frac{(c_1+c_2|\omega|)\omega}{1-e^{-\beta\omega}}=\frac{c_1}{\beta}+\frac{c_1\omega}{2}+\frac{c_2}{\beta}|\omega|+O(\omega^2).
\end{aligned}
\label{S-ohmic-expansion}
\end{align}
The apparent singularity of $\omega/(1-e^{-\beta\omega})$ at zero is removable. The strictly linear spectrum, $c_2=0$, therefore produces no real-axis non-analyticity of $S$ at finite temperature. The nonzero low-frequency curvature gives a finite jump $2c_2/\beta$ in $S'(\omega)$ and the term $(2c_2/\beta)\delta(\omega)$ in $S''(\omega)$. Here a prime denotes differentiation with respect to the argument. This bath-function property is also present in the fermionic analysis of Ref.~\cite{shi2025ultracritical}; the bosonic consequences below follow from Eq.~\eqref{S-EfES-0}. The zero-temperature limit is different: $S(\omega)=\Theta(\omega)(c_1\omega+c_2\omega^2)$ already has a slope discontinuity for $c_1>0$.

\subsubsection{Condensed phase: a kink in the occupation}
Consider an isolated positive energy $l\Omega$. As in the following gapped-bath calculation, all energy derivatives and one-sided limits are taken at fixed momentum direction, suppressing the angular argument. We write $f_\varepsilon\equiv f(\varepsilon)$ and $l\Omega^\pm=l\Omega\pm0^+$, and evaluate the overlap $\Phi^{(l)}_{{\bf0},{\bf p}}$ at $\varepsilon_{\bf p}=l\Omega$. For a small offset $\eta>0$, the paired condensate channels at $\varepsilon_{\bf p}=l\Omega+\eta$ sample $S(-\eta)$ and $S(\eta)$. Their common kink contribution to the scattering rate is $2\pi|\chi_0|^2\Phi^{(l)}_{{\bf0},{\bf p}}c_2\eta/\beta$. It enters the source $I_{\bf p}$ multiplied by $n_0$, but cancels to this order in the condensate part of $\Gamma_{\bf p}$, since
\begin{align}
\begin{aligned}
&S(\eta)-S(-\eta)=(c_1+c_2\eta)\eta,\qquad \eta>0.
\end{aligned}
\label{S-ohmic-rate-difference}
\end{align}
Thus $\Gamma$ and its first derivative have common limits. The excited-state integral does not add a slope jump: even with the condensed infrared occupation $f_{\bf q}\propto\varepsilon_{\bf q}^{-1}$, the phase-space weight $\sqrt{\varepsilon_{\bf q}}$ makes the one-sided contribution from integrating the bath kink proportional to $\eta^{3/2}$, which has a continuous first derivative at the surface. Differentiating $I=\Gamma f$ on the two sides consequently gives
\begin{align}
\begin{aligned}
&f'(l\Omega^+)-f'(l\Omega^-)=\frac{4\pi|\chi_0|^2n_0c_2\Phi^{(l)}_{{\bf0},{\bf p}}}{\beta\Gamma(l\Omega)}.
\end{aligned}
\label{S-ohmic-condensed-kink}
\end{align}
For $\Phi^{(l)}_{{\bf0},{\bf p}}>0$ and $\Gamma(l\Omega)>0$ this is a nonzero kink: $f$ is continuous, but its first derivative jumps. The surface is absent along a direction where the overlap vanishes. Its radius is $p_{F,l}=\sqrt{2ml\Omega}$ and is independent of density. This finite-temperature result requires nonzero low-frequency curvature; setting both $\nu_g$ and $c_2$ to zero is a different bath.

\subsubsection{Normal state: the band-bottom phase-space integral}
Now take $n_0=0$ and assume a finite, direction-independent band-bottom occupation $f(0)=\lim_{{\bf q}\to{\bf0}}f_{\bf q}$, a Floquet overlap that varies smoothly near ${\bf q}={\bf0}$ and $\varepsilon_{\bf p}=l\Omega$, and a finite positive $\Gamma(l\Omega)$. The two singular bath factors are equal to leading order, so their Bose-enhanced gain-minus-loss bracket reduces to $f(0)(1+f_{l\Omega})-f_{l\Omega}[1+f(0)]=f(0)-f_{l\Omega}$. At ${\bf q}={\bf0}$ the overlap is independent of the source direction, so the normalized angular integral gives unity. In three dimensions, the Dirac delta in $S''$ selects $\varepsilon_{\bf q}=\eta$ above the surface and $\varepsilon_{\bf q}=-\eta$ below it. The phase-space measure $m\sqrt{2m\varepsilon_{\bf q}}\,d\varepsilon_{\bf q}/(2\pi^2)$ therefore gives, to leading order,
\begin{align}
\begin{aligned}
&f''(l\Omega+\eta)-f''(l\Omega-\eta)\simeq\frac{2m\sqrt{2m}|\chi_0|^2c_2\Phi^{(l)}_{{\bf0},{\bf p}}}{\pi\beta\Gamma(l\Omega)}[f(0)-f_{l\Omega}]\sqrt\eta,\qquad \eta>0.
\end{aligned}
\label{S-ohmic-normal-coefficient}
\end{align}
The overlap is again evaluated at $\varepsilon_{\bf p}=l\Omega$. The square-root term is present only above the surface, since below it the delta function would require $\varepsilon_{\bf q}=-\eta<0$. The chain rule gives $\partial_\eta[f''(l\Omega+\eta)-f''(l\Omega-\eta)]=f'''(l\Omega+\eta)+f'''(l\Omega-\eta)$. At this isolated surface, the lower-side derivative has no corresponding inverse-square-root singularity. Differentiating the delta-selected collision integral therefore gives the leading upper-side term; smooth variations of its coefficient give less singular contributions:
\begin{align}
\begin{aligned}
&f'''(l\Omega+\eta)\simeq\frac{m\sqrt{2m}|\chi_0|^2c_2\Phi^{(l)}_{{\bf0},{\bf p}}}{\pi\beta\Gamma(l\Omega)}\frac{f(0)-f_{l\Omega}}{\sqrt\eta},\qquad \eta>0.
\end{aligned}
\label{S-ohmic-normal-order}
\end{align}
The second derivative has a continuous square-root kink, while the third derivative diverges on the higher-energy side. The corresponding leading non-analytic occupation contribution has a one-sided $5/2$ power. The leading coefficient vanishes if the overlap or $f(0)-f_{l\Omega}$ vanishes. Angular averaging averages the displayed coefficient and preserves the order unless that average cancels. These statements concern the finite-temperature three-dimensional normal state away from the condensation threshold.
\subsection{bosonic Floquet Fermi surfaces for the gapped bath component}
\label{S-gapped-bath-author-notation}

{The curves labeled ``gapped'' in main-text Figs.~\ref{fig2}(a,b) and \ref{fig3}(a) use the mixed bath $\nu_B(\omega)=(c_1\omega+c_2\omega^2)\Theta(\omega)+\nu_g\Theta(\omega-\Delta)$, with $c_2\Omega/c_1=\nu_g/(c_1\Omega)=1$; the gapped component is added to the gapless background.}

For the gapped bath ($\nu_g>0$), the gapped component is described by
\begin{align}
\begin{aligned}
&\nu_B^{\rm g}(\omega)=\nu_g\Theta(\omega-\Delta),\qquad \Delta>0.
\end{aligned}
\end{align}
For this contribution,
\begin{align}
\begin{aligned}
&S_{\rm g}(\omega) =\nu_g\big[N_b(|\omega|)+\Theta(\omega)\big] \Theta(|\omega|-\Delta).
\end{aligned}
\label{S-gapped-S-function}
\end{align}
Following the direct analysis of the bath spectral step in Ref.~\cite{shi2025ultracritical},
we use $[h(\omega)\Theta(\omega-a)]'=h'(\omega)\Theta(\omega-a)
+h(a)\delta(\omega-a)$ to obtain
\begin{align}
\begin{aligned}
&S_{\rm g}'(\omega)=\nu_g \begin{cases} \partial_\omega N_b(\omega)\Theta(\omega-\Delta) +[N_b(\Delta)+1]\delta(\omega-\Delta),&\omega>0,\\ \partial_\omega N_b(-\omega)\Theta(-\omega-\Delta) -N_b(\Delta)\delta(\omega+\Delta),&\omega<0. \end{cases}
\end{aligned}
\label{S-gapped-S-derivative}
\end{align}
Here a prime denotes differentiation with respect to the argument. The delta functions
give the emission and absorption spectral steps, with weights
$\nu_g[N_b(\Delta)+1]$ and $\nu_gN_b(\Delta)$, respectively.
$S_{\rm g}(\omega)=0$ for $|\omega|<\Delta$.

For the condensed three-dimensional gas, the first term of
Eq.~\eqref{S-EfES-0} evaluates the bath function at the discrete condensate
state ${\bf q}={\bf0}$. The resulting bosonic Floquet Fermi surfaces occur at
\begin{align}
\begin{aligned}
&\varepsilon_{\bf p}=l\Omega\pm\Delta.
\end{aligned}
\label{S-condensate-gap-surfaces}
\end{align}
Only positive energies $l\Omega\pm\Delta$ lie in the system spectrum. We take all
energy derivatives and one-sided limits at fixed momentum direction,
suppressing the angular argument, and write $l\Omega^\pm=l\Omega\pm0^+$. We use $f_\varepsilon\equiv f(\varepsilon)$.
The forward and reverse harmonics are paired by
$\Phi^{(-l)}_{{\bf p},{\bf0}}=\Phi^{(l)}_{{\bf0},{\bf p}}$.

At an isolated energy $l\Omega\pm\Delta$ where the condensate scattering rates change discontinuously, let $+$ and $-$ denote the two energy limits at fixed momentum direction. These are limits within the same steady state: $n_0$ and the source occupations $f_{\bf q}$ are fixed. The excited-state momentum integrals in $I_{\bf p}[f]$ and $\Gamma_{\bf p}[f]$ are continuous, because integration over ${\bf q}$ averages the scattering-rate steps over the occupied momenta. Their condensate terms need not be continuous. The definitions following Eq.~\eqref{S-EfES-0} give $\Gamma^\pm f^\pm=I^\pm$, and
\begin{align}
\begin{aligned}
&I^+-I^-=n_0\big(W_{{\bf0}\to{\bf p}}^+-W_{{\bf0}\to{\bf p}}^-\big),\qquad \Gamma^+-\Gamma^-=n_0\Big[ W_{{\bf p}\to{\bf0}}^+-W_{{\bf p}\to{\bf0}}^- -W_{{\bf0}\to{\bf p}}^++W_{{\bf0}\to{\bf p}}^-\Big].
\end{aligned}
\label{S-condensate-two-sided-balance}
\end{align}
In particular, $\Gamma^+$ and $\Gamma^-$ are generally different. Subtracting the two balance equations and rearranging the product gives the exact identities
\begin{align}
\begin{aligned}
&I^+-I^-=\Gamma^+f^+-\Gamma^-f^-=\Gamma^+(f^+-f^-)+f^-(\Gamma^+-\Gamma^-)=\Gamma^-(f^+-f^-)+f^+(\Gamma^+-\Gamma^-).
\end{aligned}
\label{S-condensate-product-identities}
\end{align}
Substituting Eq.~\eqref{S-condensate-two-sided-balance} yields two equivalent expressions,
\begin{align}
\begin{aligned}
&\Gamma^+(f^+-f^-) =n_0\Big\{(1+f^-)(W_{{\bf0}\to{\bf p}}^+-W_{{\bf0}\to{\bf p}}^-) -f^-(W_{{\bf p}\to{\bf0}}^+-W_{{\bf p}\to{\bf0}}^-)\Big\}, \\
&\Gamma^-(f^+-f^-) =n_0\Big\{(1+f^+)(W_{{\bf0}\to{\bf p}}^+-W_{{\bf0}\to{\bf p}}^-) -f^+(W_{{\bf p}\to{\bf0}}^+-W_{{\bf p}\to{\bf0}}^-)\Big\}.
\end{aligned}
\label{S-condensate-jump-identities}
\end{align}
Adding them gives the symmetric form
\begin{align}
\begin{aligned}
&(\Gamma^++\Gamma^-)(f^+-f^-) =n_0\Big\{(2+f^++f^-)(W_{{\bf0}\to{\bf p}}^+-W_{{\bf0}\to{\bf p}}^-) -(f^++f^-)(W_{{\bf p}\to{\bf0}}^+-W_{{\bf p}\to{\bf0}}^-)\Big\}.
\end{aligned}
\label{S-condensate-symmetric-jump}
\end{align}
The factors multiplying $f^-$ in the first line of Eq.~\eqref{S-condensate-jump-identities} include the change $\Gamma^+-\Gamma^-$. No equality of the two net relaxation rates, or expansion in the occupation jump, has been assumed.

The occupation jump has a simple physical origin. Just below the surface at $\varepsilon=\Omega-\Delta$, a condensate boson can absorb energy $\Omega$ from the drive and emit energy $\Omega-\varepsilon>\Delta$ into the gapped bath component. Crossing the surface from below closes this emission channel and its reverse absorption process. The change of the stationary occupation follows by subtracting the balance equations on the two sides, as in Eq.~\eqref{S-condensate-jump-identities}.

At $l\Omega-\Delta$, the condensate-to-excited-state emission channel
closes, together with its reverse absorption channel. Their rate changes
are $-2\pi|\chi_0|^2\nu_g\Phi^{(l)}_{{\bf0},{\bf p}}[N_b(\Delta)+1]$
and $-2\pi|\chi_0|^2\nu_g\Phi^{(l)}_{{\bf0},{\bf p}}N_b(\Delta)$.
At $l\Omega+\Delta$, the absorption channel and its reverse emission
channel open, giving the corresponding positive changes with the two
Bose factors exchanged. Substitution gives
\begin{align}
\begin{aligned}
&f_{l\Omega^+-\Delta}-f_{l\Omega^--\Delta} =-\frac{2\pi|\chi_0|^2n_0\nu_g\Phi^{(l)}_{{\bf0},{\bf p}}} {\Gamma(l\Omega^+-\Delta)} \Big\{[1+f_{l\Omega^--\Delta}][N_b(\Delta)+1] -f_{l\Omega^--\Delta}N_b(\Delta)\Big\}, \\
&f_{l\Omega^++\Delta}-f_{l\Omega^-+\Delta} =\frac{2\pi|\chi_0|^2n_0\nu_g\Phi^{(l)}_{{\bf0},{\bf p}}} {\Gamma(l\Omega^++\Delta)} \Big\{[1+f_{l\Omega^-+\Delta}]N_b(\Delta) -f_{l\Omega^-+\Delta}[N_b(\Delta)+1]\Big\}.
\end{aligned}
\label{S-condensate-one-sided-jumps}
\end{align}
The overlap in each line is evaluated at the corresponding energy $l\Omega-\Delta$ or $l\Omega+\Delta$.
These expressions follow from the first line of Eq.~\eqref{S-condensate-jump-identities}, with $f^-$ and $\Gamma^+$. Its second line gives the equivalent expressions with $f^+$ and $\Gamma^-$. Both retain the full change of the net relaxation rate.
For a positive net relaxation rate, the jump at $\varepsilon=l\Omega-\Delta$ is downward when $\Phi^{(l)}_{{\bf0},{\bf p}}>0$, because its bracket reduces to $N_b(\Delta)+1+f_{l\Omega^--\Delta}$. At $\varepsilon=l\Omega+\Delta$, the bracket is $N_b(\Delta)-f_{l\Omega^-+\Delta}$, so the jump can have either sign. At zero bath temperature, the first jump is downward for $\Phi^{(l)}_{{\bf0},{\bf p}}>0$, while the second is downward when also $f_{l\Omega^-+\Delta}>0$. If several channels open or close at the same energy, add their rate changes in the
subtracted balance before dividing by the full $\Gamma^+$ at that energy.

The bath gap and drive frequency fix the energies $l\Omega\pm\Delta$, while the complete
Floquet-Boltzmann equation fixes the jump amplitudes.
For the circular drive,
$\Phi^{(l)}_{{\bf0},{\bf p}}=J_l^2[A_0p\sin\theta/(m\Omega)]$.
A zero of this overlap removes the jump, including on the polarization
axis for $l\ne0$ because $J_l(0)=0$. Thus a spherical bosonic Floquet Fermi surface can have
an angle-dependent jump height.

In equilibrium, detailed balance cancels the paired emission and
absorption contributions. Explicitly,
$N_b(\omega)+1=e^{\beta\omega}N_b(\omega)$ gives
$S(-\omega)=e^{-\beta\omega}S(\omega)$.
Without driving, only $l=0$ survives. The only positive energy selected by Eq.~\eqref{S-condensate-gap-surfaces} is then $\varepsilon=\Delta$, where $f=N_b(\Delta)$ cancels the second line of Eq.~\eqref{S-condensate-one-sided-jumps}. The same cancellation occurs in an $l=0$-only kinetic problem, as proved for arbitrary pairs in SM Sec.~\ref{S-pairwise-balance}. In a driven problem with nonzero sidebands, the occupation at $\Delta$ need not equal $N_b(\Delta)$, so the $l=0$ channel can contribute a jump even though its isolated rate ratio is thermal.

For the normal gas, we self-consistently obtain $n_0=0$. The collision balance retains the
integral over ${\bf q}$, so we differentiate it with respect to
$\varepsilon_{\bf p}$, keeping $f_{\bf q}$ fixed:
\begin{align}
\begin{aligned}
&\Gamma(\varepsilon_{\bf p})f'(\varepsilon_{\bf p}) =\int[d{\bf q}]\Big[ f_{\bf q}(1+f_{\bf p})\partial_{\varepsilon_{\bf p}}W_{{\bf q}\to{\bf p}} -f_{\bf p}(1+f_{\bf q})\partial_{\varepsilon_{\bf p}}W_{{\bf p}\to{\bf q}} \Big].
\end{aligned}
\end{align}
Using $q^2dq=m\sqrt{2m\varepsilon_{\bf q}}\,d\varepsilon_{\bf q}$, the
three-dimensional phase-space measure is
\begin{align}
\begin{aligned}
&\int[d{\bf q}] =\frac{m\sqrt{2m}}{2\pi^2}\int_0^\infty d\varepsilon_{\bf q}\,\sqrt{\varepsilon_{\bf q}} \int\frac{d\Omega_{\bf q}}{4\pi}.
\end{aligned}
\end{align}
Near a positive energy $\varepsilon_i=l\Omega\pm\Delta$, the delta
terms in Eq.~\eqref{S-gapped-S-derivative} select
$\varepsilon_{\bf q}=\varepsilon_{\bf p}-\varepsilon_i$.
Evaluate just above and below, at $\varepsilon_{\bf p}=\varepsilon_i\pm\eta$
with small $\eta>0$. The required integrals are simply
\begin{align}
\begin{aligned}
&\int_0^\infty d\varepsilon_{\bf q}\,\sqrt{\varepsilon_{\bf q}}\, \delta(\varepsilon_{\bf q}-\eta)=\sqrt\eta,\qquad \int_0^\infty d\varepsilon_{\bf q}\,\sqrt{\varepsilon_{\bf q}}\, \delta(\varepsilon_{\bf q}+\eta)=0.
\end{aligned}
\end{align}
Assume a finite band-bottom occupation
$f(0)=\lim_{{\bf q}\to{\bf0}}f_{\bf q}$, a Floquet overlap that varies smoothly near ${\bf q}={\bf0}$ and $\varepsilon_{\bf p}=\varepsilon_i$, and
a finite positive $\Gamma$ at $\varepsilon_i$. To leading order, the
remaining factors can then be evaluated at ${\bf q}={\bf0}$ and
$\varepsilon_{\bf p}=\varepsilon_i$. The overlap is independent of the
direction of ${\bf q}$ there, so the angular integral gives unity.
Terms regular at $\varepsilon_i$ change by order $\eta$ between the two
sides, which is smaller than the square-root contribution.
Using $\partial_{\varepsilon_{\bf p}}S(\varepsilon_{\bf q}-\varepsilon_{\bf p}+l\Omega)
=-S'(\varepsilon_{\bf q}-\varepsilon_{\bf p}+l\Omega)$ and the paired
reverse harmonic, subtraction therefore gives, to leading order,
\begin{align}
\begin{aligned}
&f'(l\Omega+\eta-\Delta)-f'(l\Omega-\eta-\Delta) \simeq-\frac{m\sqrt{2m}|\chi_0|^2\nu_g\Phi^{(l)}_{{\bf0},{\bf p}}} {\pi\Gamma(l\Omega-\Delta)}\sqrt\eta\, \\
&\quad \Big\{f(0)[1+f_{l\Omega-\Delta}][N_b(\Delta)+1] -f_{l\Omega-\Delta}(1+f(0))N_b(\Delta)\Big\}, \\
&f'(l\Omega+\eta+\Delta)-f'(l\Omega-\eta+\Delta) \simeq\frac{m\sqrt{2m}|\chi_0|^2\nu_g\Phi^{(l)}_{{\bf0},{\bf p}}} {\pi\Gamma(l\Omega+\Delta)}\sqrt\eta\, \\
&\quad \Big\{f(0)[1+f_{l\Omega+\Delta}]N_b(\Delta) -f_{l\Omega+\Delta}(1+f(0))[N_b(\Delta)+1]\Big\}.
\end{aligned}
\label{S-normal-one-sided-derivatives}
\end{align}
Each overlap is again evaluated at its corresponding energy $\varepsilon_i$. The
prefactor follows from multiplying the rate factor
$2\pi|\chi_0|^2\nu_g$ by the phase-space factor
$m\sqrt{2m}/(2\pi^2)$ and dividing by $\Gamma$.

The square-root contribution to $df/d\varepsilon$ is present only for $\varepsilon>\varepsilon_i$. Integrating it
uses $\int_0^\eta ds\,\sqrt{s}=(2/3)\eta^{3/2}$; differentiating it uses
$d\sqrt\eta/d\eta=1/(2\sqrt\eta)$. Thus the same coefficients in
Eq.~\eqref{S-normal-one-sided-derivatives} give an occupation contribution
proportional to $(2/3)(\varepsilon-\varepsilon_i)^{3/2}$ above the surface
and a second-derivative contribution proportional to
$1/[2\sqrt{\varepsilon-\varepsilon_i}]$.
The occupation and its first derivative remain continuous, while the
second derivative has a one-sided inverse-square-root divergence.
A zero overlap or cancellation of the paired Bose factors removes this leading term. In equilibrium, $f(\varepsilon)/(1+f(\varepsilon))=e^{-\beta(\varepsilon-\mu)}$ cancels the $l=0$ bracket. 

Fig.~\ref{fig2} in the main text displays the angular average
$f_{\mathrm{exc}}(\varepsilon)=\frac12\int_{-1}^{1}du\,f(\varepsilon,u)$,
with $u=\cos\theta$. The average retains the same surface energies and
averages their direction-dependent coefficients. Panel (a) shows the
occupation jumps in the condensed gas. Panel (b) resolves the normal-state
second-derivative singularities at $\Omega\pm\Delta$, with the weaker
$\Delta$ feature shown in its leftmost inset. We solve the coupled stationary equations on a uniform energy grid and take
adjacent finite differences of the resulting occupations, without smoothing.
The nearest points to a singular energy
retain a finite-mesh dependence; their height does not represent a converged
value of the divergent derivative. The singular
coefficient depends on the steady-state occupations and can have either sign.

\subsection{Surface hierarchy and the scope of the comparison}
\label{S-surface-hierarchy}
The preceding derivations identify bosonic Floquet Fermi surfaces inherited from the bath at $l\Omega$ from its zero-transfer kink and at $l\Omega\pm\Delta$ from its spectral step. Condensate scattering or the band-bottom phase-space integral determines their order. Scattering from an existing surface in the excited occupation can generate further surfaces. To see this, let $f(\varepsilon_{\bf q})$ contain $A(\varepsilon_{\bf q}-\varepsilon_*)_+^\alpha$ near a positive source energy $\varepsilon_*$, with $\alpha\geq0$ and $x_+^\alpha\equiv x^\alpha\Theta(x)$ (a step for $\alpha=0$). The delta terms of $\partial_\omega S_{\rm g}$ select
\begin{align}
\begin{aligned}
&\varepsilon_{\bf q}=\varepsilon_{\bf p}-l\Omega\mp\Delta,\qquad \varepsilon_{\rm new}=\varepsilon_*+l\Omega\pm\Delta.
\end{aligned}
\label{S-hierarchy-transfer}
\end{align}
At $\varepsilon_*>0$ the density of states is smooth and nonzero. The singular part of $\partial f/\partial\varepsilon_{\bf p}$ then inherits the source exponent $\alpha$, and integration produces exponent $\alpha+1$ in $f$. For the emission spectral step, the coefficient multiplying the source singularity is proportional to $-[1+N_b(\Delta)+f(\varepsilon_{\rm new})]$; for absorption it is proportional to $N_b(\Delta)-f(\varepsilon_{\rm new})$. Both also contain the surface-averaged Floquet weight, the density of states, and $\nu_g/\Gamma$. This establishes the transfer whenever that coefficient is nonzero. Coincident contributions must be added before deciding their order or amplitude.

Repeated transfers produce the candidate family
\begin{align}
\begin{aligned}
&\varepsilon_{j,s}=j\Omega+s\Delta>0,\qquad p_{F,j,s}=\sqrt{2m\varepsilon_{j,s}},\qquad j,s\in\mathbb Z.
\end{aligned}
\label{S-two-integer-family}
\end{align}
Here $j\Omega$ is the net energy absorbed from the drive and $s$ counts the net signed shifts by the bath gap $\Delta$ along a scattering path. Only paths with allowed positive intermediate energies and nonzero paired coefficients generate a surface. This derives the two-integer location structure discussed in Ref.~\cite{shi2025ultracritical}, without importing its fermionic derivative orders. These additional surfaces need not have the same order as those inherited from the bath. For example, scattering from a condensed-state occupation jump produces a kink, while scattering from a normal-state surface with a $3/2$-power non-analyticity produces a surface with a $5/2$-power non-analyticity. With $\nu_g=0$, repeated transfers from the bath kink change energies only by integer multiples of $\Omega$ and do not generate an independent $\Delta$ index.

Fig.~\ref{fig2}(d) compares the strongest bosonic Floquet Fermi surfaces inherited from the bath for the three-dimensional parabolic model at $T_B>0$, with $c_2>0$. The gapless condensed entry denotes a finite jump in $\partial f/\partial\varepsilon$; its normal entry denotes a one-sided inverse-square-root divergence in $\partial^3f/\partial\varepsilon^3$. The gapped-bath condensed entry denotes a finite jump in $f$; its normal entry denotes the inverse-square-root divergence in $\partial^2f/\partial\varepsilon^2$. The derivative order is the first noncontinuous derivative, with $f$ itself counted as order zero. An occupation or derivative can remain continuous while its next derivative jumps or diverges. All entries assume $\Phi^{(l)}_{{\bf0},{\bf p}}>0$, a finite positive net relaxation rate at the surface, and no cancellation of the stated leading coefficient; the normal entries also assume finite band-bottom occupation. The angular average used in Fig.~\ref{fig2} preserves these locations and, when its coefficient is nonzero, these orders.

\begin{figure}[p]
\centering
\includegraphics[width=\textwidth]{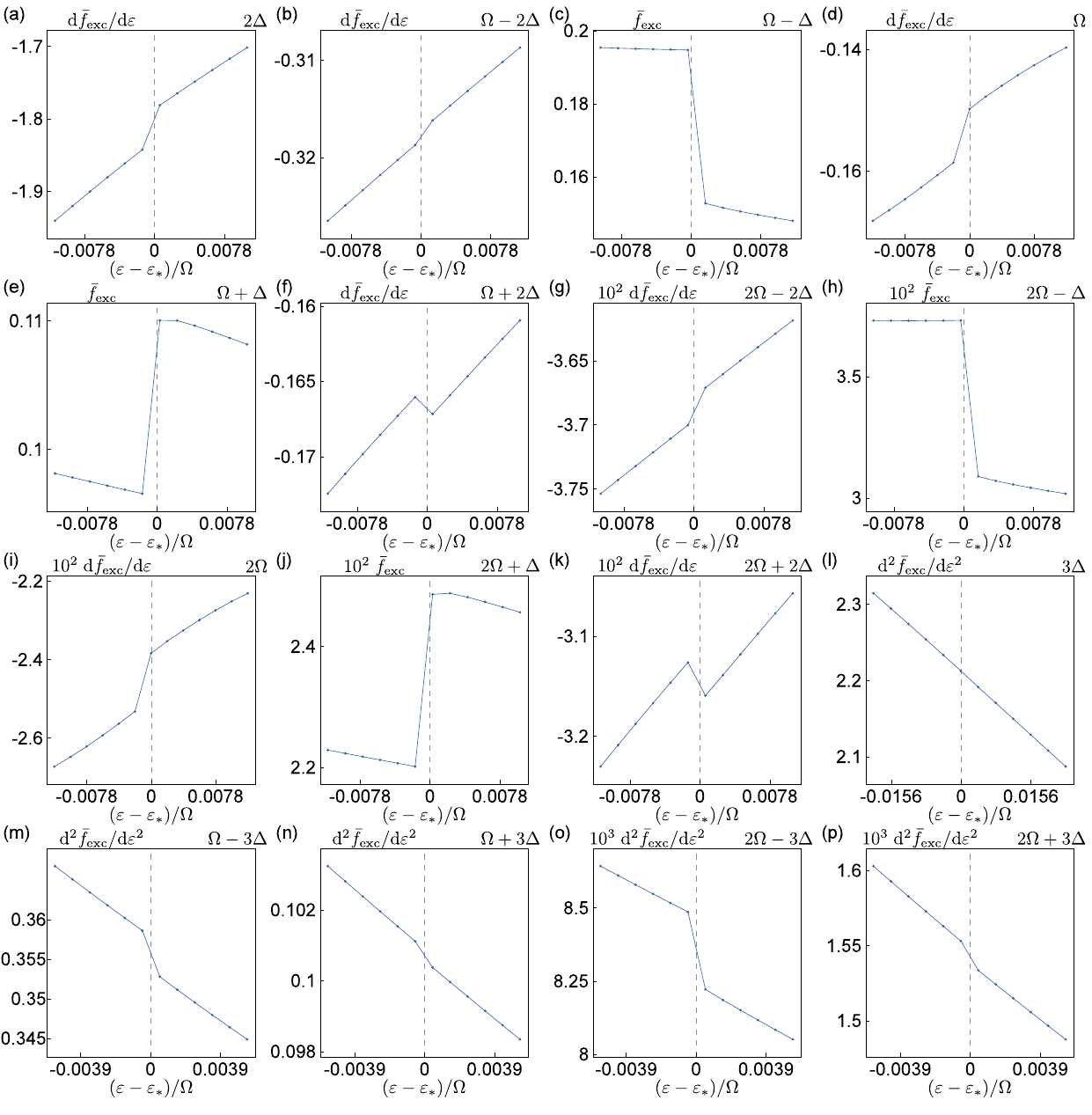}
\caption{Higher-order bosonic Floquet Fermi surfaces in the condensed gas. The angular-averaged excited occupation, denoted $\bar f_{\rm exc}$ here, and its indicated derivatives are shown near the indicated energies $\varepsilon_*=l\Omega+s\Delta>0$ for $l=0,1,2$ and $s=0,\pm1,\pm2,\pm3$. Dashed lines mark $\varepsilon_*$; dots are occupations or adjacent finite differences, connected without smoothing. Parameters are $\Delta/\Omega=\sqrt2/12$ and $(c_2\Omega/c_1,\nu_g/(c_1\Omega))=(1,1)$. The groups $(A_0/\sqrt{2m\Omega},k_BT_B/\Omega,n_{\rm tot}/(2m\Omega)^{3/2})$ are $(1,0.005,0.03)$ in (a--b,d,g,i), $(1,0.2,0.03)$ in (c,e--f,h,j--k), $(1,0.005,0.0465)$ in (l), $(0.35,0.1,0.01)$ in (m,o--p), and $(0.35,0.2,0.01)$ in (n).}
\label{figS_condensed}
\end{figure}

\begin{figure}[p]
\centering
\includegraphics[width=\textwidth]{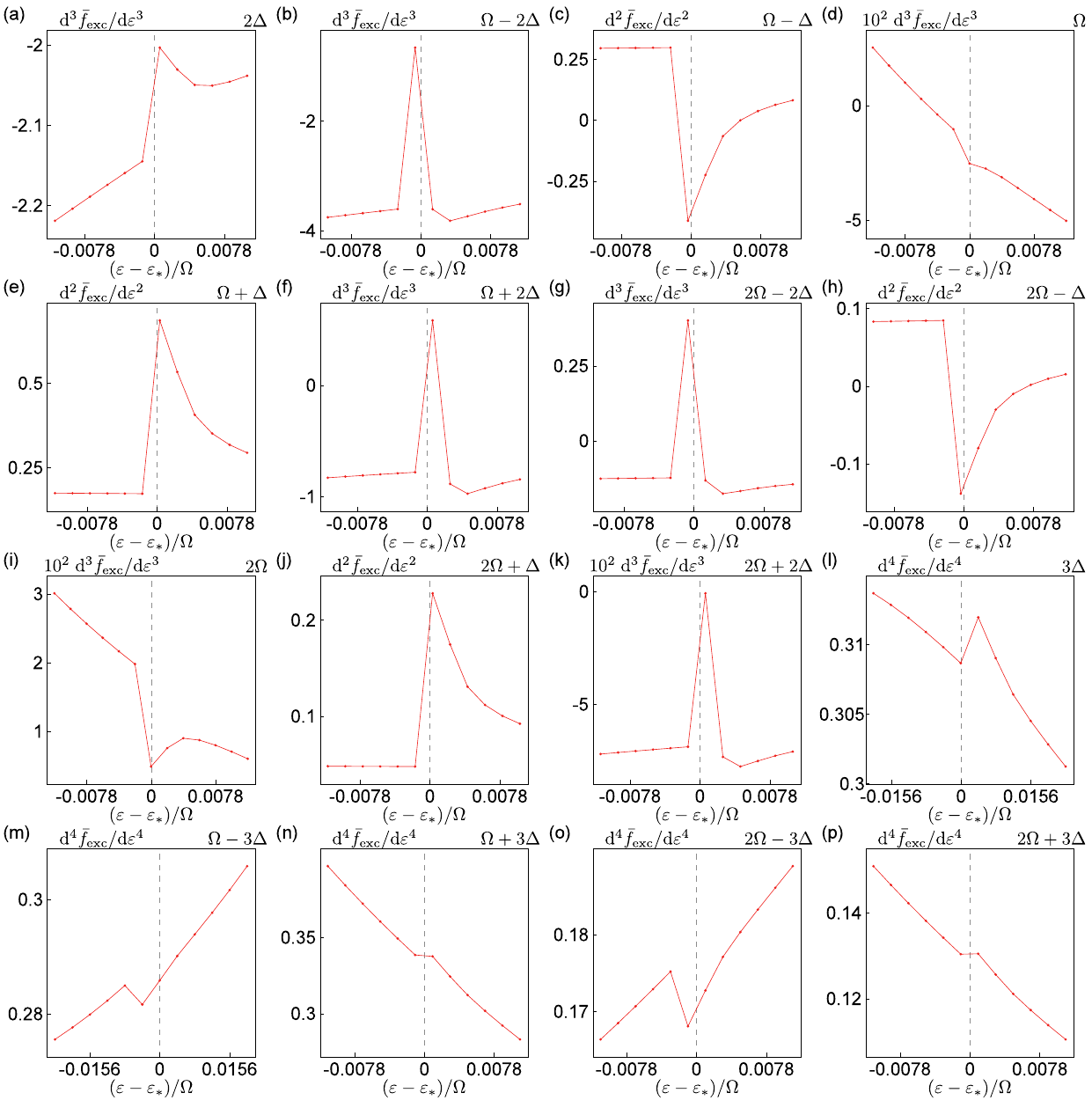}
\caption{Higher-order bosonic Floquet Fermi surfaces in the normal gas. The indicated derivatives of $\bar f_{\rm exc}$ are shown at the same sixteen energies as in Fig.~\ref{figS_condensed}. The non-analyticities occur in higher derivatives than in the condensed gas. Dashed lines mark $\varepsilon_*$; dots are adjacent finite differences of the occupations, connected without smoothing. Parameters are $\Delta/\Omega=\sqrt2/12$, $(c_2\Omega/c_1,\nu_g/(c_1\Omega))=(1,1)$, and $n_{\rm tot}/(2m\Omega)^{3/2}=0.01$. The groups $(A_0/\sqrt{2m\Omega},k_BT_B/\Omega)$ are $(1,0.005)$ in (b,f--g,k), $(1,1)$ in (a,c--e,h--j,n--p), and $(0.7,1.5)$ in (l--m).}
\label{figS_normal}
\end{figure}

Fig.~\ref{figS_condensed} and Fig.~\ref{figS_normal} show the numerical resolution of this hierarchy in the condensed and normal gases. They complement the $\Delta$ surface already shown in Fig.~\ref{fig2} with the remaining positive energies $l\Omega+s\Delta$ for the displayed range of indices. The weaker surfaces require higher derivatives to become visible. The parameter settings in the captions are chosen to resolve each surface; the different panels need not describe the same steady state.

\clearpage

\section{Exact Solutions for Driven Landau Levels and the Scattering Rates in 2D}
\setcounter{figure}{0}

The driven single-particle problem derived below is independent of quantum
statistics. After obtaining the exact Landau states, we use them as the basis
for a normal two-dimensional Bose gas. In this SM section,
$\Delta(t)$ and $\Delta_0$ denote the drive-induced scalar shift and its
cycle average; they are unrelated to the bath gap $\Delta$.

\label{S-2DLandau}

We consider a two-dimensional Bose gas in a static perpendicular synthetic magnetic field $\mathbf B=-B\hat{\mathbf z}$ and a time-periodic drive described by the vector potential $\mathbf A(t)$. We take $B>0$ for the magnetic-field magnitude; the chosen field orientation fixes its handedness relative to the drive. The single-particle Hamiltonian is given by:
\begin{align}
\begin{aligned}
&\hat h_{\rm 2D}(t) = \frac{[{\bf p} - {\bf A}_0({\bf r}) - {\bf A}(t)]^2}{2m},
\end{aligned}
\end{align}
where ${\bf A}_0({\bf r})$ generates the static magnetic field. 
With $\boldsymbol{\pi}=\mathbf p-\mathbf A_0(\mathbf r)$ and $\mathbf p=-i\nabla$, the commutator is $[\pi_x,\pi_y]=-iB$. We choose
\begin{align}
\begin{aligned}
&a=\frac{\pi_x-i\pi_y}{\sqrt{2B}}, \qquad a^\dagger=\frac{\pi_x+i\pi_y}{\sqrt{2B}}, \qquad \ell_B=\frac{1}{\sqrt{B}},
\end{aligned}
\end{align}
so that $[a,a^\dagger]=1$. Substituting
$\pi_x=\sqrt{B/2}(a+a^\dagger)$ and
$\pi_y=i\sqrt{B/2}(a-a^\dagger)$ gives the driven-oscillator Hamiltonian:
\begin{align}
\begin{aligned}
&\hat h_{\rm 2D}(t) = \omega_c\left(a^\dagger a + \frac{1}{2}\right) - a z^*(t) - a^\dagger z(t) + \frac{{\bf A}^2(t)}{2m},
\end{aligned}
\end{align}
where $\omega_c = B/m$ is the cyclotron frequency. The drive amplitude entering the term $-a^\dagger z(t)$ is defined by:
\begin{align}
\begin{aligned}
&z(t) = \sqrt{\frac{\omega_c}{2m}} [A_x(t) - iA_y(t)].
\end{aligned}
\label{S-zt}
\end{align}
For a monochromatic drive with frequency $\Omega$, we decompose $z(t)$ into its two Fourier components \begin{align}
\begin{aligned}
&z(t) = z_+ e^{-i\Omega t} + z_- e^{i\Omega t}.
\end{aligned}
\label{S-zt-1}
\end{align}

To solve the time-dependent Schrödinger equation $i \partial_t \ket{\psi(t)} = \hat h_{\rm 2D}(t) \ket{\psi(t)}$, we employ a unitary transformation using the time-dependent displacement operator $D(\alpha(t)) = \exp(\alpha(t)a^\dagger - \alpha^*(t)a)$~\cite{shi2025algebraic}. The state in the moving frame is defined as $\ket{\psi'(t)} = D^\dagger(\alpha(t)) \ket{\psi(t)}$, governed by the transformed Hamiltonian:
\begin{align}
\begin{aligned}
&\hat h_{\rm 2D}'(t) = D^\dagger(\alpha(t)) \hat h_{\rm 2D}(t) D(\alpha(t)) - i D^\dagger(\alpha(t)) \partial_t D(\alpha(t)).
\end{aligned}
\end{align}

The Baker--Campbell--Hausdorff identities needed for this transformation are
\begin{align}
\begin{aligned}
&D^\dagger aD=a+\alpha,\qquad D^\dagger a^\dagger D=a^\dagger+\alpha^*,\qquad D^\dagger\dot D =\dot\alpha\,a^\dagger-\dot\alpha^*a +\frac12(\alpha^*\dot\alpha-\alpha\dot\alpha^*).
\end{aligned}
\end{align}
For the derivative identity, write $X=\alpha a^\dagger-\alpha^*a$ and use
$e^{-X}\partial_t e^X=\dot X+[\dot X,X]/2$;
the commutator is the scalar $\alpha^*\dot\alpha-\alpha\dot\alpha^*$, so all higher commutators vanish. The coefficient of $a^\dagger$ in $\hat h'_{\rm 2D}$ is $\omega_c\alpha-z-i\dot\alpha$. We eliminate the linear terms by choosing
\begin{align}
\begin{aligned}
&\omega_c\alpha(t) - z(t) - i\dot{\alpha}(t) = 0.
\end{aligned}
\end{align}
 Away from the drive resonance, a periodic particular solution is:
\begin{align}
\begin{aligned}
&\alpha(t) = \frac{z_+}{(\omega_c - \Omega)}e^{-i\Omega t} + \frac{z_-}{(\omega_c + \Omega)}e^{i\Omega t}.
\end{aligned}
\label{S-alphat}
\end{align}
The general solution also contains $\alpha_{\rm h}e^{-i\omega_ct}$, a freely chosen cyclotron oscillation. We set this homogeneous term to zero to construct the periodic Floquet basis; the isolated driven oscillator does not relax to this choice by itself.
Under this transformation, the one-particle Hamiltonian becomes diagonal in the Landau-level basis:
\begin{align}
\begin{aligned}
&\hat h_{\rm 2D}'(t) = \omega_c\left(a^\dagger a + \frac{1}{2}\right) + \Delta(t),
\end{aligned}
\end{align}
where the time-dependent scalar shift
 \begin{align}
\begin{aligned}
&\Delta (t) = \omega_c |\alpha (t)|^2 - 2 {\rm Re}\big[ \alpha (t) z^*(t) \big] - {\rm Im} \big[ \alpha (t) \dot{\alpha}^* (t) \big] + \frac{{\bf A}^2 (t)}{2m}
\end{aligned}
\end{align}
arises from the transformation and the ${\bf A}^2(t)$ term. This shift can be decomposed into a DC component, $\Delta_0$, and an oscillating AC component, $\Delta_{\text{osc}}(t)$. 
Using the equation for $\alpha$ simplifies the scalar shift to
$\Delta(t)=\mathbf A^2(t)/(2m)-{\rm Re}[\alpha(t)z^*(t)]$.
Inserting Eq.~\eqref{S-alphat} and averaging removes the terms proportional to $e^{\pm2i\Omega t}$, giving
\begin{align}
\begin{aligned}
&\Delta_0 = -\frac{|z_+|^2}{(\omega_c-\Omega)} -\frac{|z_-|^2}{(\omega_c+\Omega)} + \left\langle \frac{{\bf A}^2(t)}{2m} \right\rangle_T.
\end{aligned}
\label{S-Delta0}
\end{align}
The exact solutions to the Schrödinger equation $i \partial_t \ket{\psi(t)} = \hat h_{\rm 2D}(t) \ket{\psi(t)}$, labeled by the Landau level index $N$, are:
\begin{align}
\begin{aligned}
&\ket{N(t)} = e^{-i\varepsilon_N t} e^{-i \Theta (t)} D(\alpha(t)) \ket{N},\qquad \varepsilon_N = \omega_c(N + 1/2) + \Delta_0
\end{aligned}
\label{eq:exact_state}
\end{align}
Here $\varepsilon_N $ represents the Floquet quasienergy of the $N$-th Landau level, and 
\begin{align}
\begin{aligned}
&\Theta(t) = \int_0^t \Delta_{\text{osc}}(t') dt',\qquad \Delta_{\text{osc}}(t) = \Delta (t) - \Delta_0.
\end{aligned}
\end{align}
This exact solution $\ket{N(t)}$ forms the basis for our subsequent analysis. The guiding-center label is unchanged by $D(\alpha)$ and is suppressed here.
\subsection{Kinetic energy and occupations of the normal Bose gas}
\label{S-B-mechanical-and-occupations}

The Floquet quasienergy in Eq.~\eqref{eq:exact_state} contains the common
shift $\Delta_0$. The quantity relevant for the energy oscillations is
instead the cycle-averaged kinetic energy. Using $D^\dagger aD=a+\alpha$, $\langle N|a|N\rangle=0$, and $\langle N|a^\dagger a|N\rangle=N$, direct evaluation in the
exact state gives
\begin{align}
\begin{aligned}
&\bar E_N \equiv\frac{1}{T}\int_0^Tdt\, \langle N(t)|\hat h_{\rm 2D}(t)|N(t)\rangle =\omega_c\left(N+\frac12\right) +\varepsilon_{\mathrm{drive}}(B),
\end{aligned}
\label{S-B-mechanical-energy}
\end{align}
where the $N$-independent part is
\begin{align}
\begin{aligned}
&\varepsilon_{\mathrm{drive}}(B)= \left\langle \omega_c|\alpha(t)|^2 -2{\rm Re}[\alpha(t)z^*(t)] +\frac{{\bf A}^2(t)}{2m} \right\rangle_T.
\end{aligned}
\label{S-B-common-drive-energy}
\end{align}
Here $\langle\cdots\rangle_T$ denotes the average over one drive period.
For the circular drive used here, $z_-=0$ and $|z_+|^2=\omega_c A_0^2/(2m)$, so
\begin{align}
\begin{aligned}
&\varepsilon_{\mathrm{drive}}(B)=\frac{A_0^2\Omega^2}{2m(\Omega-\omega_c)^2}.
\end{aligned}
\label{S-B-circular-drive-energy}
\end{align}
This kinetic-energy contribution differs from the quasienergy shift $\Delta_0=A_0^2\Omega/[2m(\Omega-\omega_c)]$. For the circular drive, $C=|\alpha|^2=\omega_c A_0^2/[2m(\Omega-\omega_c)^2]$ and $i\langle N|D^\dagger\dot D|N\rangle=\Omega C$, so
\begin{align}
\begin{aligned}
&\bar E_N=\varepsilon_N+\Omega C,\\
&\Delta_0+\Omega C=\frac{A_0^2\Omega}{2m(\Omega-\omega_c)}+\frac{A_0^2\Omega\omega_c}{2m(\Omega-\omega_c)^2}=\frac{A_0^2\Omega^2}{2m(\Omega-\omega_c)^2}.
\end{aligned}
\label{S-B-quasienergy-kinetic-relation}
\end{align}
Here $\varepsilon_N$ is the unfolded quasienergy representative in Eq.~\eqref{eq:exact_state}, whereas $\omega_c(N+1/2)$ is the undriven band energy. The correction to the former is $\Omega C$ and the correction to the latter is $\Delta_0+\Omega C$. This one-particle identity is independent of Bose or Fermi statistics. Both corrections are common to every Landau level and smooth away from cyclotron resonance. 
 
For a 2D system of area $A$, every Landau level has
\begin{align}
\begin{aligned}
&N_\phi=\frac{BA}{2\pi}
\end{aligned}
\label{S-B-Landau-degeneracy}
\end{align}
guiding-center orbitals. Their common occupation $f_N$ obeys the bosonic
Floquet-Boltzmann equation
\begin{align}
\begin{aligned}
&0=\sum_{N'=0}^{+\infty}\Big[ f_{N'}W_{N'\to N}(1+f_N) -f_NW_{N\to N'}(1+f_{N'}) \Big].
\end{aligned}
\label{S-B-normal-Landau-FBE}
\end{align}
Here $W_{N'\to N}$ includes the sum over the partner guiding-center
orbitals. For the independent local baths of SM Sec.~\ref{S-kinetics},
magnetic translation symmetry makes this rate independent of the chosen
guiding center. We may therefore average over its $N_\phi$ values.
The required sum over orbitals is the bulk Landau-level projector
\begin{align}
\begin{aligned}
&P_N(\mathbf r,\mathbf r') \equiv\sum_X\psi_{NX}(\mathbf r)\psi_{NX}^*(\mathbf r') =\frac{B}{2\pi}e^{i\varphi(\mathbf r,\mathbf r')} e^{-v/2}L_N(v),\qquad v=\frac{|\mathbf r-\mathbf r'|^2}{2\ell_B^2}.
\end{aligned}
\end{align}
Here $X$ labels the guiding center, $\psi_{NX}$ is a normalized undriven
orbital, $L_N$ is the ordinary
Laguerre polynomial, and the gauge phase satisfies
$\varphi(\mathbf r',\mathbf r)=-\varphi(\mathbf r,\mathbf r')$.
To obtain the projector, sum the Landau-gauge oscillator wavefunctions
using Mehler's formula and integrate their guiding-center coordinate.
With a generating parameter $s$, $|s|<1$, this gives
\begin{align}
\begin{aligned}
&\sum_{N=0}^\infty s^N P_N(\mathbf r,\mathbf r') =\frac{B}{2\pi}\frac{e^{i\varphi(\mathbf r,\mathbf r')}}{1-s} \exp\!\left[-\frac{1+s}{2(1-s)}v\right],\qquad |s|<1.
\end{aligned}
\end{align}
Comparison with
$\sum_{N=0}^\infty s^NL_N(v)=(1-s)^{-1}e^{-sv/(1-s)}$
gives the stated result.

For the circular drive $\mathbf A(t)=A_0(\cos\Omega t,\sin\Omega t)$
of amplitude $A_0$ in Fig.~\ref{fig3}(c,d), $z_-=0$ and we define
$C=|\alpha(t)|^2=|z_+/(\omega_c-\Omega)|^2$. The displacement $D(\alpha(t))$
translates each orbital along the circular trajectory
$\mathbf d(t)=-\sqrt2\,\ell_B({\rm Im}\alpha(t),{\rm Re}\alpha(t))$,
of radius $\sqrt{2C}\,\ell_B$, up to a phase common to all orbitals.
This phase, including $\Theta(t)$, cancels in a local-density matrix
element. Multiplying this matrix element at $t$ by its complex conjugate
at $t'$ and summing over the two guiding-center labels gives
\begin{align}
\begin{aligned}
&\frac{1}{N_\phi}\int d^2\mathbf r\, P_N\big(\mathbf r-\mathbf d(t'),\mathbf r-\mathbf d(t)\big) P_{N'}\big(\mathbf r-\mathbf d(t),\mathbf r-\mathbf d(t')\big) =\frac{B}{2\pi}e^{-v(\theta)}L_N\big(v(\theta)\big) L_{N'}\big(v(\theta)\big), \\
&v(\theta)=2C(1-\cos\theta),\qquad \theta=\Omega(t-t').
\end{aligned}
\end{align}
The projector gauge phases cancel because their arguments are reversed;
the spatial integral contributes $A$, and $N_\phi=BA/(2\pi)$ leaves
one factor $B/(2\pi)$. Here $v(\theta)=|\mathbf d(t)-\mathbf d(t')|^2/(2\ell_B^2)$
is the squared separation along the orbit in magnetic-length units. Rotational symmetry makes the remaining overlap
depend only on $\theta$. The dimensionless weight for a scattering channel exchanging energy $l\Omega$ with the drive is its Fourier coefficient,
\begin{align}
\begin{aligned}
&\Phi_{NN'}^{(l)}(C)=\int_0^{2\pi}\frac{d\theta}{2\pi}\, \cos(l\theta)e^{-v(\theta)}L_N\big(v(\theta)\big)L_{N'}\big(v(\theta)\big).
\end{aligned}
\label{S-B-circular-time-weight}
\end{align}
Its nonnegativity follows from the sum of squared $l$-th Fourier
components of the local-density matrix element.
The scattering rate is
\begin{align}
\begin{aligned}
&W_{N'\to N}=2\pi|\chi_0|^2\frac{B}{2\pi} \sum_{l=-\infty}^{\infty}\Phi_{NN'}^{(l)}(C) S\big((N'-N)\omega_c+l\Omega\big).
\end{aligned}
\label{S-B-circular-Landau-rates}
\end{align}
The common quasienergy shift $\Delta_0$ cancels from the transition
energy. The guiding-center factor is already included in $W$; no further
factor is needed in Eq.~\eqref{S-B-normal-Landau-FBE}. The numerical rate matrix divides all rates by the common factor $B/(2\pi)$ at each field. This leaves the stationary occupations unchanged; restoring that factor recovers the physical rates and collision balances.
At zero drive, $v=0$ and $L_N(0)=1$, so
$\Phi_{NN'}^{(l)}(0)=\delta_{l0}$.

The sum of Eq.~\eqref{S-B-normal-Landau-FBE} over $N$ vanishes, because the
bath redistributes bosons without changing their total number. Bose
stimulation is already contained in $1+f_N$ and does not introduce another
degeneracy. Thus
\begin{align}
\begin{aligned}
&n_{\mathrm{tot}} =\frac{B}{2\pi}\sum_{N=0}^{+\infty}f_N,\qquad \frac{\bar E}{A} =\frac{B}{2\pi}\sum_{N=0}^{+\infty} \omega_c\left(N+\frac12\right)f_N +n_{\mathrm{tot}}\varepsilon_{\mathrm{drive}}(B).
\end{aligned}
\label{S-B-particle-energy-sums}
\end{align}
At fixed
$n_{\mathrm{tot}}$, the last term is a common smooth background; the oscillatory part comes from the sum over Landau-level energies $\omega_c(N+1/2)$.

\subsection{Gapless-bath surfaces and inverse-field periods}
\label{S-B-gapless-periods}
For the gapless bath with $c_2>0$, the finite-temperature kink in $S(\omega)$ at $\omega=0$ produces occupation non-analyticities at $\varepsilon_l=l\Omega>0$. Their two-dimensional zero-field areas are $\mathcal S_l=\pi p_{F,l}^2=2\pi ml\Omega$, and the corresponding inverse-field periods are
\begin{align}
&\Delta(1/B)_l=\frac{2\pi}{\mathcal S_l}=\frac{1}{ml\Omega}.
\label{S-B-gapless-period}
\end{align}
The same commensurabilities appear directly in Eq.~\eqref{S-B-circular-Landau-rates}: a channel crosses the bath kink when $(N'-N)\omega_c+l\Omega=0$. Consecutive values of $N-N'$ therefore differ by $1/(ml\Omega)$ in inverse field. The $l=1$ channels give the main dips in Fig.~\ref{fig3}(c); the $l=2$ channels allow additional features halfway between them. At strong drive the occupations are determined from all retained channels together. A comparison of separately solved sideband truncations identifies this mechanism, but their difference is not an additive independent two-photon contribution to the energy.

\subsection{Quantum oscillations in the kinetic energy and the equilibrium Fermi reference}
\label{S-fig3-background}
At fixed density, the Landau-level contribution to the kinetic energy per particle is
\begin{align}
&\bar\varepsilon_{\rm band}(B)=\frac{B}{2\pi n_{\rm tot}}\sum_N\omega_c(N+1/2)f_N,
\qquad\bar\varepsilon_{\rm kin}(B)=\bar\varepsilon_{\rm band}(B)+\varepsilon_{\rm drive}(B).
\label{S-B-mean-kinetic-energy}
\end{align}
Since $[B/(2\pi n_{\rm tot})]\sum_N f_N=1$, the field-dependent occupation and degeneracy do not change the common per-particle drive correction. We remove the smooth background $\bar\varepsilon_{\rm sm}(B)$ of the band energy and include the exact drive term in the corresponding kinetic-energy background:
\begin{align}
&\delta\bar\varepsilon_{\rm kin}(B)=\bar\varepsilon_{\rm kin}(B)-[\bar\varepsilon_{\rm sm}(B)+\varepsilon_{\rm drive}(B)]
=\bar\varepsilon_{\rm band}(B)-\bar\varepsilon_{\rm sm}(B).
\label{S-B-kinetic-residual}
\end{align}
For the Bose curves, least-squares separation of a smooth polynomial background in $B$ and harmonics of $m\Omega/B$ with smoothly varying amplitudes determines $\bar\varepsilon_{\rm sm}$; only this background is subtracted from the calculated energy. No phenomenological broadening is applied.

The reference in Fig.~\ref{fig3}(d) is an equilibrium, single-component ideal Fermi gas with no Zeeman splitting and $A_0=0$. Its occupations are the exact Fermi--Dirac values
\begin{align}
&f_N^{\rm F}=\frac{1}{\exp\{[\omega_c(N+1/2)-\mu]/(k_BT)\}+1},\qquad
n_{\rm tot}=\frac{B}{2\pi}\sum_N f_N^{\rm F},
\label{S-B-Fermi-reference}
\end{align}
with $\mu$ chosen at each field and temperature to maintain the same density as the Bose gas.

For each gas, the oscillation magnitude is half the peak-to-peak variation over the common inverse-field interval $\mathcal I$ specified in Fig.~\ref{fig3}:
\begin{align}
&\mathcal A_{\rm B,F}(T)=\frac12\left[\max_{1/B\in\mathcal I}\delta\bar\varepsilon_{\rm kin}(B,T)-\min_{1/B\in\mathcal I}\delta\bar\varepsilon_{\rm kin}(B,T)\right].
\label{S-B-temperature-amplitude}
\end{align}
This gives the usual amplitude for a sinusoid when the interval contains both extrema, and retains all harmonics and the field-dependent envelope of the calculated curves. 

\section{Density Correlations and bosonic Floquet Fermi Surfaces}
\label{S-C-exact}
\setcounter{figure}{0}

We relate the equal-time density correlations of the three-dimensional gas to its steady-state occupation and determine the oscillations produced by the bosonic Floquet Fermi surfaces. We then specify the correlation evaluation and background subtraction used in Fig.~\ref{fig3} and Fig.~\ref{figS2}. Throughout this section the magnetic field is zero, $\varepsilon_{\bf p}=p^2/(2m)$, and $\hbar=1$.

\subsection{Density correlations from the occupation}
\label{S-C-formalism}
\label{S-C-bunching}

For separated points, define the connected density correlation and the normalized pair correlation by
\begin{align}
\begin{aligned}
&C_B({\bf r})=\langle\hat n({\bf0})\hat n({\bf r})\rangle-n_{\mathrm{tot}}^2,
\qquad g_B({\bf r})=\frac{\langle\hat n({\bf0})\hat n({\bf r})\rangle}{n_{\mathrm{tot}}^2},
\\
&C_B({\bf r})=n_{\mathrm{tot}}^2[g_B({\bf r})-1],\qquad {\bf r}\ne{\bf0}.
\end{aligned}
\label{S-C-definitions}
\end{align}
Thus $C_B$ is the connected correlator in the main text, while $g_B-1$ is its dimensionless normalization.

Separate the condensate from the excited states, $\hat\psi({\bf r})=\sqrt{n_0}+\delta\hat\psi({\bf r})$. The excited-state one-body density matrix is
\begin{align}
\begin{aligned}
&\tilde f_{\mathrm{exc}}({\bf r})=\langle\delta\hat\psi^\dagger({\bf0})\delta\hat\psi({\bf r})\rangle
=\int\frac{d^3{\bf p}}{(2\pi)^3}e^{i{\bf p}\cdot{\bf r}}f_{\bf p},
\qquad \tilde f_{\mathrm{exc}}({\bf0})=n_{\mathrm{exc}},
\qquad n_{\mathrm{tot}}=n_0+n_{\mathrm{exc}}.
\end{aligned}
\label{S-C-onebody}
\end{align}
For the uniform drive considered here, the time-dependent phase of each Floquet plane wave cancels between the bra and ket. Equation~\eqref{S-C-onebody} therefore holds at each drive time. Inversion symmetry, $f_{-{\bf p}}=f_{\bf p}$, makes $\tilde f_{\mathrm{exc}}$ real.

As in SM Sec.~\ref{FFGR}, we evaluate the excited-state four-operator averages from pair averages, with $\langle\delta\hat\psi\,\delta\hat\psi\rangle=0$. The condensate has no thermal bunching term: its relative number fluctuations vanish in the thermodynamic limit. This is the condensate correction discussed in Sec.~III B of Ref.~\cite{NaraschewskiGlauber1999}. At separated points, the two contractions of the excited fields and their interference with the condensate give
\begin{align}
\begin{aligned}
&\langle\hat n({\bf0})\hat n({\bf r})\rangle
=n_0^2+2n_0n_{\mathrm{exc}}+n_{\mathrm{exc}}^2
+2n_0\operatorname{Re}\tilde f_{\mathrm{exc}}({\bf r})+|\tilde f_{\mathrm{exc}}({\bf r})|^2,
\\
&C_B({\bf r})=2n_0\operatorname{Re}\tilde f_{\mathrm{exc}}({\bf r})+|\tilde f_{\mathrm{exc}}({\bf r})|^2.
\end{aligned}
\label{S-C-pair}
\end{align}
The quadratic term is the bunching of excited particles. The linear term is their interference with the condensate; it enhances the surface oscillations at large separation, where $\tilde f_{\mathrm{exc}}$ is small. At coincident points we use the normally ordered pair correlation, excluding the contact term $n_{\mathrm{tot}}\delta({\bf r})$, which gives $g_B^{(2)}({\bf0})=2-(n_0/n_{\mathrm{tot}})^2$.

\subsection{Surface contributions at large separation}
\label{S-C-driven}
\label{S-C-driven-surfaces}

Consider first an angle-independent occupation jump $A_i\Theta(p_{F,i}-p)$, where $A_i=f(p_{F,i}-0^+)-f(p_{F,i}+0^+)$ and $p_{F,i}=\sqrt{2m\varepsilon_i}$. Its Fourier transform is
\begin{align}
\begin{aligned}
&\tilde f_{\mathrm{jump},i}({\bf r})
=\frac{A_i}{2\pi^2r}\int_0^{p_{F,i}}dp\,p\sin(pr)
=\frac{A_i}{2\pi^2r^3}[\sin(p_{F,i}r)-p_{F,i}r\cos(p_{F,i}r)].
\end{aligned}
\label{S-C-single-step}
\end{align}
The leading term oscillates at $p_{F,i}$ with an $r^{-2}$ envelope. A smooth angular dependence of the jump changes its amplitude. At large $r$, the angular integral is dominated by momenta parallel and antiparallel to $\hat{\bf r}$, giving
\begin{align}
\begin{aligned}
&\tilde f_{\mathrm{jump}}({\bf r})\simeq
-\frac{1}{4\pi^2r^2}\sum_i p_{F,i}
\left[A_i(\hat{\bf r})e^{ip_{F,i}r}+A_i(-\hat{\bf r})e^{-ip_{F,i}r}\right]
=-\frac{1}{2\pi^2r^2}\sum_i p_{F,i}A_i(\hat{\bf r})\cos(p_{F,i}r),
\end{aligned}
\label{S-C-directional-steps}
\end{align}
where the last equality uses inversion symmetry. In the condensed gas, the gapped component of the bath produces jumps at $\varepsilon_i=l\Omega\pm\Delta>0$. Inserting Eq.~\eqref{S-C-directional-steps} into the linear term of Eq.~\eqref{S-C-pair} gives their leading oscillatory contribution,
\begin{align}
\begin{aligned}
&[C_B({\bf r})]_{\mathrm{osc}}
\simeq-\frac{n_0}{\pi^2r^2}\sum_i p_{F,i}A_i(\hat{\bf r})\cos(p_{F,i}r).
\end{aligned}
\label{S-C-condensed-oscillations}
\end{align}
The radii are fixed by the drive and bath, whereas the amplitudes depend on the solved state and the observation direction. The oscillatory subscript selects the finite-radius surface contribution; Eq.~\eqref{S-C-pair} itself contains the full correlation.

Weaker occupation non-analyticities can be treated in the same way. Let a local surface contribution be $B(\hat{\bf p})(p-p_{F,i})^\gamma\Theta(p-p_{F,i})$, with $\gamma>-1$ and a smooth cutoff away from the surface. Its radial endpoint integral contributes $r^{-1-\gamma}$ and the angular integral contributes $r^{-1}$. For inversion-symmetric $B$,
\begin{align}
\begin{aligned}
&\tilde f_{\mathrm{surf}}({\bf r})\simeq
\frac{p_{F,i}\Gamma(1+\gamma)B(\hat{\bf r})}{2\pi^2r^{2+\gamma}}
\sin\!\left[p_{F,i}r+\frac{\pi}{2}(1+\gamma)\right],
\end{aligned}
\label{S-C-edge-symmetric}
\end{align}
where $\Gamma$ is the Euler gamma function. Since $\varepsilon-\varepsilon_i=(p_{F,i}/m)(p-p_{F,i})+O[(p-p_{F,i})^2]$, the radial-momentum and energy exponents agree. In particular, a kink in the occupation has $\gamma=1$ and gives an $r^{-3}$ condensate-interference contribution. This is the case at $\varepsilon_i=l\Omega>0$ for the finite-temperature gapless bath with $c_2\ne0$.

\subsubsection{The normal gas}
\label{S-C-driven-normal}

Without a condensate, $C_B=|\tilde f_{\mathrm{exc}}|^2$. The three-dimensional normal-state surfaces inherited from the gapped bath component have $\gamma=3/2$ when the coefficient in Eq.~\eqref{S-normal-one-sided-derivatives} is nonzero. Their one-body contribution consequently has an $r^{-7/2}$ envelope, and its square has an $r^{-7}$ envelope. The latter includes oscillations at $2p_{F,i}$ and $p_{F,i}\pm p_{F,j}$, as well as a nonoscillatory part. The normal-state gapless-bath surface with $\gamma=5/2$ instead gives a one-body $r^{-9/2}$ contribution. These statements describe the specified surface contributions; cross terms with other parts of the occupation are also present in the full correlation. Condensation enhances the spatial signature through the linear term in Eq.~\eqref{S-C-pair}, but is not required for surface-induced oscillations.

For comparison, the ideal three-dimensional single-component Fermi gas at zero temperature has $g_F-1=-|\tilde f|^2/n_{\mathrm{tot}}^2$ and $n_{\mathrm{tot}}=p_F^3/(6\pi^2)$. Equation~\eqref{S-C-single-step} with $A=1$ therefore gives
\begin{align}
\begin{aligned}
&g_F(r)-1=-9\frac{[\sin(p_Fr)-p_Fr\cos(p_Fr)]^2}{(p_Fr)^6}
\simeq-\frac{9\cos^2(p_Fr)}{(p_Fr)^4}.
\end{aligned}
\label{S-C-fermion-comparison}
\end{align}
The $r^{-2}$ oscillation of the condensed Bose gas follows from condensate interference, while the fermionic pair correlation is quadratic in the one-body function.

\subsection{Numerical correlations and real-space background subtraction}
\label{S-C-driven-background}

We evaluate the density correlations from the complete steady-state occupation and condensate density using Eq.~\eqref{S-C-pair}. The angular integral is expressed in spherical Bessel functions, and the remaining radial integral is evaluated numerically. This retains both the surface oscillations and the nonoscillatory background.

For the condensed gas, the leading large-distance nonoscillatory background in a fixed direction has the form (for its equilibrium counterpart, see Problem~12.9, p.~304, of Ref.~\cite{huang1987statistical} and Eqs.~(13.6.40)--(13.6.41), p.~525, of Ref.~\cite{pathria2011statistical})
\begin{align}
\begin{aligned}
&[g_B(r\hat{\bf r})-1]_{\rm nonosc}\simeq\frac{C_g(\hat{\bf r})}{r},\qquad r\to\infty,\\
&C_g(\hat{\bf r})=\lim_{r\to\infty}r[g_B(r\hat{\bf r})-1].
\end{aligned}
\label{S-C-CgD-def}
\end{align}
For Fig.~\ref{fig3}(a,b), we estimate this single coefficient by least squares from the real-space correlation over a fixed long-distance interval $\mathcal I$, and subtract only its $1/r$ contribution:
\begin{align}
\begin{aligned}
&\hat C_g=\underset{c}{\operatorname{argmin}}\sum_{r_j\in\mathcal I}\left[g_B(r_j\hat{\bf r})-1-\frac{c}{r_j}\right]^2,\\
&\delta g_B(r)=g_B(r\hat{\bf r})-1-\frac{\hat C_g}{r}.
\end{aligned}
\label{S-C-real-space-subtraction}
\end{align}
This procedure requires neither the momentum distribution nor the condensate density. The residual retains faster-decaying nonoscillatory terms and the surface oscillations, without multiplication by powers of $r$. Varying the fitting interval checks their stability. A coefficient error $\delta C_g=\hat C_g-C_g$ leaves a smooth contribution $-\delta C_g/r$ to the residual.

\begin{figure}[t]
\centering
\includegraphics[width=105mm]{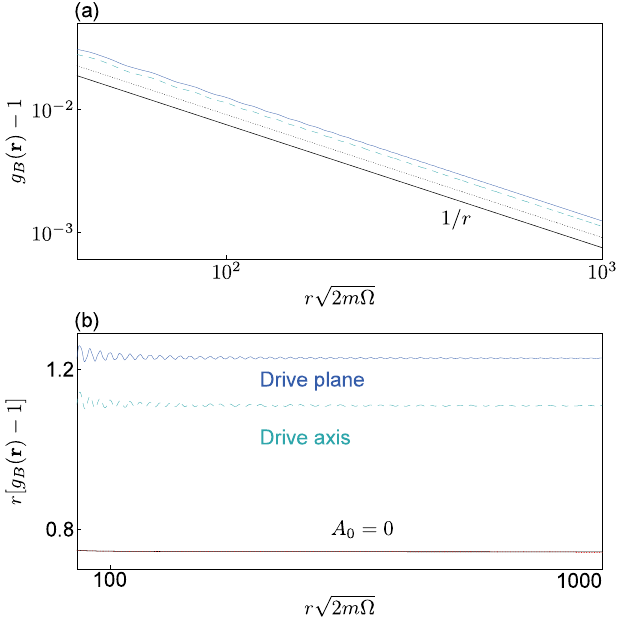}
\caption{Density correlations of the three-dimensional condensed gas at $k_BT_B/\Omega=0.05$.
(a) The normalized connected correlation $g_B({\bf r})-1=C_B({\bf r})/n_{\mathrm{tot}}^2$ along the drive plane (blue) and drive axis (teal), with a $1/r$ guide (black).
(b) Multiplication by $r$ displays the nonoscillatory background and the surface oscillations. The undriven numerical curve (black) is compared with the exact equilibrium result (red dotted), obtained from Eq.~(13.6.34), p.~524, of Ref.~\cite{pathria2011statistical} and Eq.~\eqref{S-C-pair}.
Other parameters follow Fig.~\ref{fig2}(a); the undriven comparison sets $A_0=0$ at the same bath temperature and total density.}
\label{figS2}
\end{figure}

The undriven comparison uses the same total density and bath temperature. Its exact excited-state one-body function is obtained from Eq.~(13.6.34) of Ref.~\cite{pathria2011statistical} and inserted into Eq.~\eqref{S-C-pair}. This provides an equilibrium comparison for the full numerical correlations in Fig.~\ref{figS2}.

\end{document}